\documentclass[review, 3p, times, 12pt, authoryear]{elsarticle}

\usepackage[utf8]{inputenc}
\usepackage[T1]{fontenc}

\usepackage{amsmath}
\usepackage{amsfonts}
\usepackage{amssymb}

\usepackage{float}
\usepackage[normalem]{ulem}

\usepackage{tikz}
\usepackage{pgfplots}
\pgfplotsset{compat=1.16}

\usepackage{threeparttable}
\usepackage{booktabs}
\usepackage{dcolumn}
\usepackage{longtable}
\usepackage{tabulary}
\usepackage{threeparttablex}
\usepackage{tabularx}
\usepackage{ltablex}
\usepackage{multirow}

\newcolumntype{M}{>{\raggedright\arraybackslash}X}
\usepackage{makecell}

\usepackage{siunitx}
\DeclareSIUnit{\wattth}{Wth}
\DeclareSIUnit{\wattel}{We}
\DeclareSIUnit{\year}{a}
\DeclareSIUnit{\wattpeak}{Wp}
\DeclareSIUnit{\million}{million}
\DeclareSIUnit{\coo}{\ce{CO2}e}

\usepackage[gen]{eurosym}
\usepackage[version=4]{mhchem}
\defcitealias{BMNT2018}{BMNT and BMVIT, 2018}
\defcitealias{Regierungsprogramm2020}{BKA, 2020}
\defcitealias{WSB2019}{WSB, 2019}
\defcitealias{ENTSOE2020a}{ENTSO-E, 2020a}
\defcitealias{ENTSOE2020b}{ENTSO-E, 2020b}
\defcitealias{AGEB2018}{AGEB, 2018}
\defcitealias{CCCS2017}{C3S, 2017}
\defcitealias{IMF2020}{IMF (2020)}
\defcitealias{opsd2019}{OPSD (2019)}

\usepackage{graphicx}
\usepackage{caption}
\usepackage{subcaption}

\usepackage{etoolbox}
\patchcmd{\MaketitleBox}{\footnotesize\itshape\elsaddress\par\vskip36pt}{\footnotesize\itshape\elsaddress\par\parbox[b][36pt]{\linewidth}{\vfill\hfill\textnormal{\today}\hfill\null\vfill}}{}{}
\patchcmd{\pprintMaketitle}{\footnotesize\itshape\elsaddress\par\vskip36pt}{\footnotesize\itshape\elsaddress\par\parbox[b][36pt]{\linewidth}{\vfill\hfill\textnormal{\today}\hfill\null\vfill}}{}{}

\usepackage[hidelinks, colorlinks=false, citecolor=blue, linkcolor=blue, filecolor=blue,urlcolor=blue,pdfauthor=Sebastian Wehrle]{hyperref}

\newcommand*{\figref}[2][]{%
  \hyperref[{fig:#2}]{%
    Figure~\ref*{fig:#2}%
    \ifx\\#1\\%
    \else
      \,#1%
    \fi
  }%
}

\makeatletter
\def\ps@pprintTitle{%
 \let\@oddhead\@empty
 \let\@evenhead\@empty
 \def\@oddfoot{}%
 \let\@evenfoot\@oddfoot}
\makeatother

\begin{document}
    \begin{frontmatter}
        \title{The Cost of Undisturbed Landscapes}
        \author[1]{Sebastian Wehrle\corref{cor1}}
        \ead{sebastian.wehrle@boku.ac.at}
        \author[1]{Johannes Schmidt}
        \ead{johannes.schmidt@boku.ac.at}
        \cortext[cor1]{Corresponding author}
        \address[1]{Institute for Sustainable Economic Development, University of Natural Resources and Life Sciences,
        Feistmantelstrasse 4, 1180 Vienna, Austria}

        \begin{abstract}
            By 2030 Austria aims to meet \SI{100}{\percent} of its electricity demand from domestic renewable sources, with most of the additional generation coming from wind and solar energy.
            Apart from the benefit of reducing \ce{CO2} emissions and, potentially, system cost, wind power is associated with negative impacts at the local level, such as interference with landscape aesthetics.
            Some of these impacts might be avoided by using alternative renewable energy technologies.
            Thus, we quantify the opportunity cost of wind power versus its best feasible alternative solar photovoltaics, using the power system model \emph{medea}.
            Our findings suggest that the cost of undisturbed landscapes is considerable, particularly when solar PV is mainly realized on roof-tops.
            Under a wide range of assumptions, the opportunity cost of wind power is high enough to allow for significant compensation of the ones affected by local, negative wind turbine externalities.
        \end{abstract}

    \end{frontmatter}
    \newpage
\section{Introduction}\label{sec:introduction}
    "Holding the increase in the global average temperature [\ldots] well below \SI{2}{\celsius} above pre-industrial levels and pursuing efforts to limit the temperature increase to \SI{1.5}{\celsius} above pre-industrial levels"~\citep{UNFCCC2015} requires rapid and substantial reductions in greenhouse gas emissions worldwide~\citep{IPCC2018}.
    Increasing the share of electricity generated from renewable sources consequently is amongst the United Nations' sustainable development goals (indicator 7.2.1) and a prominent goal of many climate and energy strategies.
    The European Union, for example, aims to meet at least \SI{32}{\percent} of its final energy consumption from renewable sources by 2030~\citep{EU2018}.\footnote{In its "European Green Deal" the EU Commission proposes to increase ambitions and reduce \ce{CO2} emissions by at least \SI{55}{\percent} by 2030. This compares to a \SI{40}{\percent} reduction target previously and is likely to have knock-on effects on the expansion of renewable energy generation.}
    Against this background, Austria aims to generate \SI{100}{\percent} of its electricity consumption from domestic renewable sources on annual balance by 2030\footnote{Industry's own use and ancillary service provision are exempt.    Together, these accounted for approximately \SI{10}{\percent} of consumption in 2016.}.
    The required expansion of renewable energy generation shall be incentivized by technology-specific subsidies rather than \ce{CO2} pricing.
    
    According to government estimates, an additional \SI{27}{\tera\watt\hour} of electricity needs to be generated from renewable sources annually to meet this policy goal.
    Out of the targeted total increase in renewable electricity generation, \SI{1}{\tera\watt\hour} is thought to be sourced from biomass, and \SI{5}{\tera\watt\hour} from new hydropower generation, according to the government's climate and energy strategy~\citepalias{BMNT2018}.
    The comparatively small projected increase in electricity generation from these sources results from the perception of biomass being ecologically\footnote{More specifically, increasing biomass harvests would decrease carbon stocks strongly, both through deforestation and management. Thus, a net positive greenhouse gas balance of bioenergy would only be possible after a rather long payback time. However, additions to the stock of atmospheric greenhouse gases need to be reduced rapidly.}
    ~\citep{Erb2018} and economically unsustainable\footnote{The temporary expiration of Austria's subsidy scheme for biomass plants in 2019 resulted in a widespread shutdown of biomass-fired generators.}
    , while further development of Austria's hydropower potential was considered `challenging'~\citep{Wagner2015} even before the projected \SI{5}{\tera\watt\hour} increase in hydro generation was factored in.
    As Austrians banned the use of nuclear power in a referendum in 1978~\citep{Pelinka1983}, wind and solar power are the only remaining options for a large-scale expansion of low-carbon power generation in the country.

    While wind power mitigates negative external effects at the global (\ce{CO2} emissions) and the regional (air pollutants from fossil thermal power generation) level, it also imposes negative external effects at the local level.
    Wind turbines are found to impact wildlife negatively~\citep{Loss2013,Voigt2015,Wang2015} and to emit noise~\citep{Wang2015a}.
    They are also associated with land-use and land-cover change~\citep{Turkovska2020} and with negative interference with landscape aesthetics in general~\citep{Jones2010, Meyerhoff2010}.
    The latter is found to be the dominant, negative external effect of wind turbines~\citep{Mattmann2016}.
    Accordingly, several strands of literature seek to quantify the negative impact of wind turbines, for example through their effect on property prices within a revealed preference framework~\citep{Droees2016, Gibbons2015,
    Heintzelmann2017, Jensen2018, Kussel2019, Lang2014, Sims2008, Sunak2016, Vyn2014}, or through choice experiments and surveys~\citep{Drechsler2011, Mattmann2016, Meyerhoff2010} in a stated preferences-context.
    \cite{Zerrahn2017} provides a comprehensive survey of the literature on wind power externalities.

    Apart from negative external effects at the local level, wind power is also found to have the potential to reduce electricity system costs, in particular, relative to solar PV.
    For the case of Germany, \cite{Ueckerdt2013} estimate solar PV integration cost to be about four times higher than the integration cost of wind power at penetration levels of \SI{25}{\percent}. 
    \cite{Scholz2017} find integration cost of \SI[per-mode=symbol,sticky-per, bracket-unit-denominator=false]{40}[\euro]{\per\mega\watt\hour} for a mix of \SI{80}{\percent} solar PV and \SI{20}{\percent} wind power at \SI{100}{\percent} renewables penetration in Europe.
    This reduces to below \SI[per-mode=symbol,sticky-per, bracket-unit-denominator=false]{25}[\euro]{\per\mega\watt\hour} when the technology mix consists of \SI{20}{\percent} solar PV and \SI{80}{\percent} wind power.
    These findings illustrate that abandoning wind power for PV in favor of undisturbed landscapes potentially comes at significant opportunity cost.
    We complement the literature on negative wind turbine impacts and set out to quantify this cost of undisturbed landscapes for the case of Austria.
    First, Austrian policies targeting a fixed share of renewable electricity generation are an ideal framework for studying the effects of substituting renewable energy generation technologies (RET).
    Moreover, Austria has large hydro storage capacities (including seasonal hydro storages) in place, which should limit PV integration cost.
    Thus, our findings should be qualitatively generalizable to power systems with high renewable shares but lower storage capacities with climates comparable to Austria.
    Methodologically, we rely on the power system model \emph{medea}, which is described in \autoref{sec:data-and-methods}.

    Our analysis contributes to a more comprehensive understanding of the trade-offs faced in the design of highly renewable electricity systems, thereby helping to inform policies about the socially optimal expansion of RET.
    Any such policy necessarily needs to account for the full social cost, including all externalities, of RET.
    Further, we contribute to the analysis of near-optimal power system configurations~\citep[see e.g.][]{Neumann2019, Schlachtberger2017}.
    Corresponding results are presented in \autoref{sec:results-discussion}.
    Finally, our investigation into the cost of undisturbed landscapes generates insight into the possibilities and limitations of funding potential compensating measures for the benefit of the ones negatively affected by wind turbines. 
    We illustrate these potentials by an approximation of the valuation of undisturbed landscapes that is implicit in Austria's climate and energy strategy.
    \hyperref[sec:conclusions-policy-implication]{Section~\ref*{sec:conclusions-policy-implication}} summarizes policy implications and concludes.


\section{Data and Methods} \label{sec:data-and-methods}
    To investigate the cost of undisturbed landscapes (i.e. the opportunity cost of wind power) we use the power system model \emph{medea}, as summarized in \autoref{subsec:medea}.
    We instantiate the model with scenario assumptions and data as observed in 2016, which was a low-wind year in Austria, while solar and hydro resources were close to long-term averages (see \autoref{subsec:data}).
    Based on this, we approximate the opportunity cost of wind power through the following procedure:
    \begin{enumerate}
        \item We derive the unrestricted system cost-minimizing deployment of wind and solar power, given the scenario assumptions
        \item We restrict deployment of wind power by a small margin (so that PV, the next best RET, substitutes for wind power) and observe net system cost $c_{net}$ for Austria, calculated as total system cost including air pollution cost net of the balance of trade.\label{enumerate:restrict}
        \item We repeat step~\ref{enumerate:restrict} till no wind power can be deployed.
        \item Finally, we approximate the opportunity cost of wind power (at given wind power capacity $w$) $OC_w$ by the change in $c_{net}$ in response to a change in wind power capacity $w$ deployed, i.e.\ \[OC_w = \frac{\Delta c_{net}}{\Delta w}\]
        where $\Delta$ is the difference operator and opportunity cost $OC_w$ are expressed in \EUR per \SI{}{\mega\watt} wind power foregone.
    \end{enumerate}

\subsection{Power system model \emph{medea}} \label{subsec:medea}
    We make use of the power system model \emph{medea} to simulate (dis)investment in, and hourly operation of, the prospective Austrian and German power systems in the year 2030.
    The model is cast as a linear optimization seeking to minimize total system cost, which consists of fuel and emission costs, quasi-fixed and variable operation and maintenance (O\&M) costs, the costs of investment in energy generation, storage, and transmission assets, and the (potential) cost of non-served load.
    From an economic perspective, the model reflects a perfectly competitive energy-only market with fully price-inelastic demand and perfect foresight of all actors.

    The modeled system is required to meet exogenous and inelastic demand for electricity and heat at any hour of the year.
    Energy supply, in turn, is constrained by available installed capacities of energy conversion, storage, and transmission units.
    Co-generation units convert fuel to heat and power subject to a feasible operating region defined by the unit's electrical efficiency, the electricity loss per unit of heat production, and the backpressure coefficient.
    Electricity generation from intermittent sources (wind, run-of-river hydro, solar) is subject to exogenous hourly generation profiles, which are scaled according to total installed capacities.
    Electricity from these sources can be curtailed at no additional cost (free disposal).

    Electricity can be stored in reservoir and pumped hydro storages, batteries, or hydrogen storage (more precisely: water electrolysis, hydrogen storage, and reconversion to electricity in fuel cells).
    The capacity of hydro storages can not be expanded, as we assume existing potentials to be exhausted.
    Battery and hydrogen storage capacities, on the other hand, can be added endogenously.
    Generation from storage is constrained by installed capacity and stored energy.
    Inflows of water into reservoirs add to stored energy.
    Pumped hydro storages, batteries, and hydrogen storage can actively store electricity for later use.
    To better capture operational differences of hydro storage units, we model daily, weekly and seasonal reservoir and pumped storage plants separately.
    To ensure the stable and secure operation of the electricity system, ancillary services (e. g. frequency control, voltage support) are required.
    We model ancillary service needs as a minimum requirement on spinning reserves operating at any point in time.
    Thus, we assume that ancillary services can be provided by thermal power plants, run-of-river hydro plants, or any (active) storage technology.
    We don't model transmission and distribution grids within Austria or Germany.
    We do, however, allow for cross-border electricity trade between both countries.
    We also keep track of the cost of (non-\ce{CO2}) air pollution arising from burning fossil fuels.
    As these costs are external, they do not enter cost minimization.
    For a detailed mathematical description of the model, please refer to \ref{sec:medea-desc}.
    Data processing is implemented in python, while the optimization model is based on GAMS. The model code is published at \url{https://github.com/inwe-boku/medea} under an open MIT license.
    Running the model in hourly resolution for one year takes 10-15 minutes, depending on the model parameters, on an Intel i7-8700 machine with 16 GB RAM, using CPLEX 12.10 as a solver.

\subsection{Data} \label{subsec:data}
    We set up our model with the goal to resemble Austria's prospective electricity and district heating systems in the year 2030.
    We also include Austria's largest electricity trading partner, Germany, to account for potential effects from electricity trade.
    For Germany, our scenario reflects current and announced future electricity sector policies up to 2030, so that we set generation capacities to levels consistent with these policies.
    These assumptions are laid out in subsections \ref{subsubsec:assumptions-austria} and \ref{subsubsec:assumptions-germany}, and summarized in \autoref{tab:instcaps}.

\subsubsection{Scenario Assumptions for Austria} \label{subsubsec:assumptions-austria}
    The Austrian government has set itself the goal of generating \SI{100}{\percent} of electricity consumption from domestic renewable sources on annual balance by 2030~\citepalias{Regierungsprogramm2020}.
    However, industry own consumption and system services, which currently account for about \SI{10}{\percent} of annual electricity consumption, are exempt.
    The government plans to achieve this goal by generating an additional \SI{27}{\tera\watt\hour} of electricity annually from renewable sources\footnote{Please note that this policy goal does not imply that the generated electricity must actually be consumed.   Hence, we count curtailed electricity as contributing to the policy goal.}
    , for a total of \SI{78.1}{\tera\watt\hour} of renewable electricity generation in 2030.
    New hydropower plants are thought to contribute \SI{5}{\tera\watt\hour} annually, while additional electricity generation from biomass is envisaged to account for \SI{1}{\tera\watt\hour} annually.
    In our scenarios, we add generation capacities sufficient to reach these targets.
    Further, the government projects the remainder to come from solar PV and onshore wind turbines with an annual contribution of \SI{11}{\tera\watt\hour} and \SI{10}{\tera\watt\hour} \citepalias{Regierungsprogramm2020}, respectively.
    
    As we are interested in determining the opportunity cost of wind power versus its best alternative, we allow for endogenous investment in wind and solar power without enforcing announced technology-specific targets.
    Other low-carbon energy technologies are not feasible at a large scale in the Austrian context, as we have laid out in the introduction.
    Initial generation capacities for Austria are summarized in \autoref{tab:instcaps}.

\subsubsection{Scenario Assumptions for Germany} \label{subsubsec:assumptions-germany}
    Germany has announced specific capacity targets for several power generation technologies.
    Following these announcements, we anticipate an end to nuclear power generation, a (partial) coal exit according to recommendations by the `coal commission'~\citepalias{WSB2019}, and a further expansion of renewable electricity generation in line with the German Renewable Energy Sources Act~\citep{noauthor_renewable_2017}, except for the \SI{52}{\giga\watt} cap on solar PV, which was removed meanwhile.
    Investment in additional capacity (as well as decommissioning of excess capacities) is incentivised through technology-specific government subsidies, so that we assume targeted capacities will be in place by 2030.
    The corresponding capacity assumptions are displayed in \autoref{tab:instcaps}.
    In addition, we allow for endogenous (dis-) investment in energy conversion capacities.
\begin{ThreePartTable}
    \centering
    \renewcommand\TPTminimum{0.8\textwidth}
    \begin{TableNotes}
        \footnotesize
        \item[a] capacity scaled to match generation from energy balance \citep{StatistikAustria2020} 
    \end{TableNotes}
    \begin{longtable}{l c c c c}
        \caption{Initial Energy Conversion Capacities (2030)} \label{tab:instcaps}\\
        \toprule
        Technology       & \multicolumn{2}{c}{Austria} & \multicolumn{2}{c}{Germany}\\
        \cmidrule(lr){2-3}\cmidrule(lr){4-5}
                         & \si{\giga\wattel} & \si{\giga\wattth} & \si{\giga\wattel} & \si{\giga\wattth}\\
        \midrule
        \endhead
        \midrule
        \multicolumn{5}{c}{\textit{continued on next page}}     \endfoot
        \bottomrule
        \insertTableNotes
        \endlastfoot
        Wind Onshore     & $2.6$ & --    & $90.8$ & --    \\
        Wind Offshore    & --    & --    & $15.0$ & --    \\
        Solar PV         & $1.1$ & --    & $73.0$ & --    \\
        Run-of-river Hydro & $6.8$ & --  & $4.5$  & --    \\
        Hydro storage    & $8.7$ & --    & $6.5$  & --    \\
        Biomass          & $0.7$ & $0.4$ & $8.4$  & $4.9$ \\
        Lignite          & --    & --    & $11.4$ & $8.8$ \\
        Coal             & --    & --    & $14.0$ & $9.9$ \\
        Natural Gas      & $3.9$ & $1.7$ & $24.2$ & $6.1$ \\
        Mineral Oil      & $0.2$ & --    & $3.5$  & $1.7$ \\
        Heat Pump        & --    & $1.0$ & --     & $1.0$ \\
        Gas Boiler       & --    & $1.5$ & --     & $18.0$\\
        Biomass Boiler   & --    & $0.8$\tnote{a} & --     & 2.6     \\ \bottomrule
    \end{longtable}
\end{ThreePartTable}

\subsubsection{Energy supply}
    We represent 21 energy conversion and storage technologies that are expected to be operated in 2030.
    In addition to given, initially installed capacities (see \autoref{tab:instcaps}), the model can endogenously add further generation capacities that are compatible with stated policy objectives.
    We calculate the annualized investment cost of each technology based on an assumed weighted average cost of capital of \SI{5}{\percent} over the plant's lifetime.
    All technologies can also be decommissioned so that we adopt a long-run perspective on the power system.
    \autoref{itm:techparms} summarizes the parameters of admissible technologies.
\begin{ThreePartTable}
    \renewcommand\TPTminimum{\textwidth}
        \begin{TableNotes}
            \footnotesize
            \item SC -- subcritical, USC -- (ultra-)supercritical, ST -- steam turbine, GT -- gas turbine, CC -- combined cycle, PSP -- Pumped storage plant
            \item[a] Alkaline electrolysis + cavern storage + PEM fuel cell
            \item[b] own assumption
            \item[c] \EUR / $(\text{MVA} \times \text{km})$
            \item Sources: 
            [1] \cite{DEA2019}, 
            [2] \cite{Schroeder2013},
            [3] \cite{LacalArantegui2014},
            [4] \cite{OekoInstitut2017},
            [5] \cite{DEA2019},
            [6] \cite{DEA2020a},
            [7] \cite{Brown2018},
            [8] \cite{Hagspiel2014}
        \end{TableNotes}
        
        \begin{longtable}{l ccc ccc l}
        \caption{Technology Input Parameters for 2030} \label{itm:techparms}\\
        \toprule
        &\multicolumn{2}{c}{Capital Cost} & \multicolumn{2}{c}{O\&M-Cost} & & & \\ \cmidrule(lr){2-3}\cmidrule(lr){4-5}
        Technology & Power & Energy & Quasi- & Variable & Life- & Efficiency & Source   \\
        & & & fixed & & time & &          \\
        &\EUR/kW & \EUR/kWh & \EUR/MWa & \EUR/MWh & a & & \\
        \midrule
        \endhead
        \midrule
        \multicolumn{8}{c}{\textit{continued on next page}} \endfoot
        \bottomrule
        \insertTableNotes
        \endlastfoot
        Wind Onshore         & $1040$ & NA    & $12600$ & $1.35$ & $30$ & NA & [1]\\
        Wind Offshore        & $1930$ & NA    & $36053$ & $2.70$ & $30$ & NA & [1]\\
        PV Rooftop           & $870$  & NA    & $10815$ & $0.00$ & $40$ & NA & [1]\\
        PV Open-space        & $380$  & NA    & $7250$  & $0.00$ & $40$ & NA & [1]\\
        Hydro run-of-river   & --     & NA    & $60000$ & $0.00$ & $60$ & NA & [2, 3] \\
        \midrule
        Lignite Adv          & --     & NA    & $40500$ & $0.85$ & $40$ & 0.439 & [4]\\
        Coal SC              & --     & NA    & $25000$ & $6.00$ & $25$ & 0.390 & [2] \\
        Coal USC             & --     & NA    & $31500$ & $3.00$ & $25$ & 0.460 & [1]\\
        Nat Gas ST           & $400$  & NA    & $15000$ & $3.00$ & $30$ & 0.413 & [2]\\
        Nat Gas GT           & $435$  & NA    & $7745$  & $4.50$ & $25$ & 0.410 & [1]\\
        Nat Gas CC           & $830$  & NA    & $27800$ & $4.20$ & $25$ & 0.580 & [1]\\
        Oil ST               & $400$  & NA    & $6000$  & $3.00$ & $30$ & 0.410 & [2]\\
        Oil GT               & $363$  & NA    & $7745$  & $4.50$ & $25$ & 0.400 & [1] \\
        Biomass              & $3300$ & NA    & $96000$ & $4.60$ & $25$ & 0.270 & [1]\\ \midrule
        Hydro Reservoir      & --     & --    & $22000$ & $3.00$ & $60$ & $0.900$\tnote{$\dagger$} & [3]\\
        Hydro PSP            & --     & --    & $22000$ & $3.00$ & $60$ & $0.810$\tnote{$\dagger$} & [3]\\
        Battery Li-Ion       & $320$ & $302$ & $540$ & $1.80$ & $25$ & 0.920 & [1]\\ 
        Hydrogen Storage\tnote{a} & $2575$ & $2$ & 82500 & 0 & 25 & 0.333 & [1, 5, 6] \\ \midrule
        Electric Boiler      & $140$  & NA    & $1020$  & $0.50$ & $20$ & 0.990 & [5]\\
        Absorption heat pump & $510$  & NA    & $2000$  & $1.30$ & $25$ & 1.730 & [5]\\
        Natural gas boiler   & $50$   & NA    & $1900$  & $1.00$ & $25$ & 0.920\tnote{b} & [5] \\ \midrule
        Transmission (AC)    & $455$\tnote{c} & NA    & $9$\tnote{c} & $0.00$ & $40$ & 1.000 & [7, 8]\\
        \end{longtable}
    \end{ThreePartTable}

    Power plants running on fossil fuels emit \ce{CO2} into the atmosphere. 
    We approximate the amount of \ce{CO2} released through the carbon intensity of fossil fuels, as displayed in \autoref{table:carbon-intensity}. 
    These estimates are based on an analysis by the German environmental agency~\citep{Juhrich2016}.
    
    In the baseline scenario, we assume capital cost of solar PV of \SI[per-mode=symbol,sticky-per, bracket-unit-denominator=false]{625}[\euro]{\per\kilo\wattpeak} installed.
    These costs are capacity-weighted average cost of rooftop and open-space PV installations, with \SI{50}{\percent} of PV capacity being mounted on rooftops, while the remainder is realized as open-space solar PV at utility-scale.
    In \autoref{subsec:sensitivity}, we analyze the sensitivity of our results to the capital cost of solar PV\@.
    
\begin{table}[t]
    \centering
    \begin{threeparttable}
        \renewcommand\TPTminimum{0.8\textwidth}
        \caption{\ce{CO2} intensity of fossil fuels in tonnes \ce{CO2} per MWh}
        \label{table:carbon-intensity}
        \begin{tabular}{l c c c c c}
            \toprule
            &  & Lignite & Coal & Natural Gas & Mineral Oil  \\
            \midrule
            \ce{CO2} intensity & \si[per-mode=symbol,sticky-per, bracket-unit-denominator=false]{\tonne\per\mega\watt\hour} & 0.399 & 0.337 & 0.201 & 0.266        \\
            \bottomrule
        \end{tabular}
        \begin{tablenotes}
            \footnotesize
            \item Source: \cite{Juhrich2016}
        \end{tablenotes}
    \end{threeparttable}
\end{table}

    Generation from non-dispatchable technologies solar PV, wind turbines, and run-of-river hydro is assumed to follow hourly generation profiles\footnote{A \emph{generation profile} tracks the share of installed intermittent capacity that is generating electricity over time.} 
    as observed in 2016.
    German solar and wind power profiles are sourced from \cite{opsd2019}.
    For Austria, solar and wind power generation profiles are based on solar and wind power generation time series reported by \cite{opsd2019}, which we scaled to match annual generation as reported in the national energy balance~\citep{StatistikAustria2020}. 
    Subsequently, the scaled generation time series were divided by installed capacities as published in~\cite{Biermayr2019}.
    Similarly, generation profiles for run-of-river hydropower were derived from generation data reported by \cite{ENTSOE2020b}, which we scaled to match annual run-of-river electricity generation as reported by Austria's regulatory body \cite{EControl2020}.

    Data on inflows into Austrian hydro reservoirs are not publicly available.
    However, the \cite{ENTSOE2020a} reports weekly data on hydro storage filling levels along with hourly electricity generation from hydro reservoirs and pumping by pumped hydro storages \citepalias{ENTSOE2020b}.
    Based on this data, we have approximated inflows as the part of the change in (weekly) hydro storage fill levels that isn't explained by upsampled hourly pumping and generation.
    To arrive at hourly inflows, we have interpolated our weekly estimates by piecewise cubic Hermite interpolation.
    Although inflows to water reservoirs are given exogenously, all storage technologies (including hydro reservoir and pumped storage plants) are charged and dispatched endogenously.
    \autoref{tab:time-series} summarizes the descriptive statistics of the abovementioned renewables time series.

\subsubsection{Energy demand} \label{subsec:energy-demand}
    By combining government policy targets with information about electricity generation in our base year~\citep{StatistikAustria2020}, we can infer that an electricity demand of \SI{83}{\tera\watt\hour} is expected in 2030.
    To generate a time series of hourly electricity demand, we scale Austria's electricity load as reported by \cite{opsd2019} accordingly.
    For Germany, we scale reported hourly load to match annual electricity consumption as reported in national energy balances~\citepalias{AGEB2018} for the year 2016. 
    For 2030 we assume electricity demand to remain unchanged from 2016 levels.

    Annual heat consumption is derived from energy balances for Austria and Germany, respectively~\citep{AGEB2018, StatistikAustria2020}.
    Subsequently, we break annual heat consumption down into hourly heat consumption, based on standard natural gas load profiles for space heating in the residential and commercial sectors~\citep{Almbauer2008}.
    These load profiles make use of daily average temperatures to calculate daily heat demand.
    We extract spatially resolved temperature data from ERA-5 climate data sets~\citepalias{CCCS2017} and compute a capacity-weighted average of temperatures at locations of combined heat and power (CHP) generation units.
    Daily heat demand based on these weighted average temperatures is then broken down to hourly consumption based on standardized factors accounting for weekday and time-of-day effects.
    Descriptive statistics of electricity and heat consumption are provided in \autoref{tab:time-series}.
    \begin{table*}[t]
    \centering
    \begin{threeparttable}
        \caption{Descriptive statistics of time series (as observed in 2016)}\label{tab:time-series}
        \begin{tabular}{l c c c c c c l}
            \toprule
            Name & Area & Unit & mean & median & max & min & Source   \\
            \midrule
            Electricity load & AT & GW & 7.15 & 7.14 & 10.5 & 4.21 & [1]\\
            Electricity load & DE & GW & 60.2 & 59.6 & 82.8 & 34.6 &[1]\\
            District heating load\tnote{a} & AT & GW & 2.56 & 2.43 & 5.77 & 0.91 & [2, 3]\\
            District heating load\tnote{a} & DE & GW & 14.6 & 14.1 & 26.0 & 8.9 & [2, 4]\\
            \midrule
            Wind onshore profile & AT &\%   & 0.226 & 0.140 & 0.913 & 0.000 & [1, 3]         \\
            Wind onshore profile & DE &\%   & 0.176 & 0.133 & 0.764 & 0.003 & [1]\\
            Wind offshore profile & DE &\%   & 0.371 & 0.327 & 0.900 & 0.000 & [1]\\
            Solar PV profile\tnote{a} & AT &\%   & 0.114 & 0.014 & 0.666 & 0.000 & [1, 3, 5] \\
            Solar PV profile & DE &\%   & 0.099 & 0.003 & 0.661 & 0.000 & [1] \\
            Run-of-river profile\tnote{a} & AT &\%    & 0.585 & 0.559 & 0.971 & 0.202 & [6, 7]\\
            Run-of-river profile & DE &\%   & 0.437 & 0.420 & 0.637 & 0.208 & [7] \\
            \bottomrule
        \end{tabular}
        \begin{tablenotes}
            \footnotesize
            \item[a] own calculation based on referenced sources
            \item Sources: 
            [1] OPSD \citeyearpar{opsd2019}, 
            [2] C3S \citeyearpar{CCCS2017}, 
            [3] \cite{StatistikAustria2020}, 
            [4] AGEB \citeyearpar{AGEB2018}, 
            [5] \cite{Biermayr2019}, 
            [6] ENTSO-E \citeyearpar{ENTSOE2020b}, 
            [7] \cite{EControl2020}
        \end{tablenotes}
    \end{threeparttable}
    \end{table*}
\subsubsection{Electricity Transmission}
    Electricity transmission between the modeled market areas is limited to \SI{4.9}{\giga\watt}, in line with the introduction of a congestion management scheme by German and Austrian authorities in 2018~\citep{BNetzA2017}.
    Section \ref{subsubsec:sens_transmission} explores the effects of increasing transmission capacity to \SI{10}{\giga\watt}.
    
\subsubsection{Prices}
    Monthly prices for exchange-traded fuels (hard coal, crude oil (Brent), natural gas) are retrieved from the International Monetary Fund's Commodity Data Portal.
    We convert these prices to \SI[per-mode=symbol,sticky-per, bracket-unit-denominator=false]{}[\euro]{\per\mega\watt\hour} based on the fuel's energy content and market exchange rates obtained from the \cite{ECB2020}.
    Finally, we resample prices to hourly frequency using piecewise cubic Hermite interpolation.
    Due to its low energy density, lignite is not transported over large distances and consequently also not traded on markets.
    Instead, lignite-fired power plants are situated in proximity to lignite mines.
    According to estimates from~\cite{OekoInstitut2017}, the price of lignite in Germany is close to \SI[per-mode=symbol,sticky-per, bracket-unit-denominator=false]{1.50}[\euro]{\per\mega\watt\hour}.
    Biomass-fired power plants run on a wide variety of solid and gaseous fuels, some of which are marketed.
    However, the continued operation of biomass-fired plants typically relies on sufficient subsidies.
    As a first-order approximation to more complex subsidy schemes, we assume subsidized fuel cost of \SI[per-mode=symbol,sticky-per, bracket-unit-denominator=false]{6.50}[\euro]{\per\mega\watt\hour}.
    In consequence, biomass-fired plants are operating at or close to capacity throughout the scenarios considered.
    \footnote{Subsidies are not included in system cost, as the required level is uncertain. Biomass capacity can not be expanded and is constant across scenarios. Thus, electricity generation from biomass and the corresponding subsidies vary by no more than \SI{1.75}{\percent}.}
    \begin{table*}[th]
        \centering
        \begin{threeparttable}
        \caption{Descriptive data of price time series (as observed in 2016)}
        \begin{tabular}{l c r r r r c}
            \toprule
            Name & Unit & mean & median & max & min & Source                  \\
            \midrule
            Lignite &\EUR/MWh & 1.50 & 1.50 & 1.50 & 1.50 & [1] \\
            Coal &\EUR/MWh & 8.58 & 8.13 & 11.90 & 6.69 & [2] \\
            Natural gas &\EUR/MWh & 13.62 & 12.87 & 20.18 & 12.04 & [2]     \\
            Mineral oil &\EUR/MWh & 26.40 & 26.63 & 33.44 & 18.34 & [2] \\
            Biomass (subsidized) &\EUR/MWh & 6.50 & 6.50 & 6.50 & 6.50 & [3] \\
            \bottomrule
        \end{tabular}
        \begin{tablenotes}
            \footnotesize
            \item Sources:
            [1] \cite{OekoInstitut2017}, 
            [2] \citetalias{IMF2020},
            [3] own assumption
        \end{tablenotes}
        \end{threeparttable}
    \end{table*}
    Future \ce{CO2} prices impact renewables deployment ~\citep{Brown2020, Kirchner2019} and affect the results of our analysis.
    As the future efficient price of \ce{CO2} is highly uncertain, we have conducted our analysis for \ce{CO2} prices in the range of \euro$0$ to \euro$100$ per tonne.
    Implicitly, we assume efficient pricing of \ce{CO2} emissions, i.e.\ all otherwise external cost of \ce{CO2} emissions are internalized through the prevailing \ce{CO2} price.

\subsubsection{The cost of air pollution}
    In addition to \ce{CO2} emissions, we also keep track of the external cost imposed by the emission of air pollutants such as nitrogen oxides, sulfur oxide, particulate matters, carbon monoxide, or heavy metals from fossil power plants. 
    For the valuation of these external costs, we rely on assessments from the NEEDS project, as reported in \cite{Samadi2017}.
    We have, however, converted these output-related estimates to input-related values, so that variable cost refer to the energy content of fuel used to generate electricity. 
    The air pollution cost of RET mainly reflect the cost of air pollution during manufacturing of the plants.
    To avoid distortions from different underlying assumptions on RET utilisation, we have converted these cost estimates to reflect the specific annual cost per unit of capacity installed.
    The corresponding estimates for the external cost of air pollution from fossil fuel combustion are summarized in \autoref{tab:air_pollution_cost}.
    \begin{table*}[ht]
        \centering
        \begin{threeparttable}
        \caption{The external cost of air pollution}
        \label{tab:air_pollution_cost}
        \centering
        \begin{tabular}{l c c}
        \toprule
            & Unit & Cost \\
        \midrule
             Lignite & \SI[per-mode=symbol,sticky-per, bracket-unit-denominator=false]{}[\euro]{\per\mega\watt\hour} & $4.06$ \\
             Coal & \SI[per-mode=symbol,sticky-per, bracket-unit-denominator=false]{}[\euro]{\per\mega\watt\hour} &  $6.12$ \\
             Natural Gas & \SI[per-mode=symbol,sticky-per, bracket-unit-denominator=false]{}[\euro]{\per\mega\watt\hour} & $2.36$\\
             Mineral Oil & \SI[per-mode=symbol,sticky-per, bracket-unit-denominator=false]{}[\euro]{\per\mega\watt\hour} & $3.21$\\
             Biomass & \SI[per-mode=symbol,sticky-per, bracket-unit-denominator=false]{}[\euro]{\per\mega\watt\hour} & $4.04$ \\
             Wind power & \SI[per-mode=symbol,sticky-per, bracket-unit-denominator=false]{}[\euro]{\per\mega\watt} & $2\,831$ \\
             Solar Photovoltaics & \SI[per-mode=symbol,sticky-per, bracket-unit-denominator=false]{}[\euro]{\per\mega\wattpeak} & $5\,028$ \\
         \bottomrule
        \end{tabular}
        \begin{tablenotes}
            \footnotesize
            \item Source: \cite{Samadi2017}, own calculations
        \end{tablenotes}

        \end{threeparttable}
    \end{table*}
    Please note that these cost are considered external and consequently do not enter system cost minimization.

    The data retrieval and processing scripts are published on GitHub \url{https://github.com/inwe-boku/medea} under the MIT license.

\section{Results and Discussion} \label{sec:results-discussion}
    To derive the social opportunity cost of not using wind power (but solar photovoltaics instead) in 2030, we start by determining the long-run power market equilibrium without constraints on renewable capacity addition.
    This gives us our unconstrained baseline scenario with minimal system costs.
    Subsequently, we report the system effects of a gradual restriction of wind power and the corresponding estimates of the cost of undisturbed landscapes (in terms of the opportunity cost of not installing wind turbines).

\subsection{Unconstrained Baseline} \label{subsec:baseline}
    By 2030, Austria generates the bulk of its electricity from intermittent sources. 
    Given our baseline assumptions and carbon price of \SI[per-mode=symbol,sticky-per, bracket-unit-denominator=false]{25}[\euro]{\per\tonne\coo}, this comes at a total system cost of  \SI{2.56}[\euro]{} billion.
    \autoref{tab:cost_baseline} summarizes the breakdown of the total system cost.
    
    Due to the large share of electricity generation from renewable sources, operation and maintenance cost of \SI{1}[\euro]{} billion exceeds spending on fuels and emission certificates required by thermal power plants.
    Burning fossil fuels gives rise to \SI{7.1}{\mega\tonne} of \ce{CO2} emissions, which are priced at \SI{177}[\euro]{} million.
    
    Thermal generation capacity, running on natural gas, waste, and biomass, is not expanded in the baseline scenario. 
    Hence, the annualized investment cost is entirely spent on the expansion of renewable electricity generation capacity. 
\begin{table*}[th]
    \centering
    \begin{threeparttable}
        \caption{System cost components and trade balance, baseline scenario (2030, \SI[per-mode=symbol,sticky-per, bracket-unit-denominator=false]{25}[\euro]{\per\tonne\coo})}
        \label{tab:cost_baseline}
        \begin{tabular}{l c c c c c}
        \toprule
            & Unit & Fuel \& \ce{CO2} &  Investment & O\&M & Trade Balance\\
            \midrule
            Total Cost & million \euro & $824.2$ & $1019.3$ & $718.2$ & $-699.0$\\
            \bottomrule
        \end{tabular}
        \begin{tablenotes}
            \footnotesize
            \item Source: own calculations
        \end{tablenotes}
    \end{threeparttable}
\end{table*}
    In our baseline scenario, \SI{10.6}{\giga\watt} of onshore wind turbines are added to the power plant stock, while no investment in solar PV or storage technologies, such as batteries and hydrogen storage, occurs. 
    In total, \SI{59.7}{\tera\watt\hour} electricity is generated from intermittent renewable sources, with another \SI{5.7}{\tera\watt\hour} generated by biomass-fired units and \SI{9.4}{\tera\watt\hour} from inflows into hydro storage reservoirs for a total of \SI{74.8}{\tera\watt\hour} of electricity generated from domestic renewable sources.
    To reduce quasi-fixed and variable operation and maintenance cost, electricity generation units with a capacity of \SI{0.3}{\giga\watt} (natural gas-fired) and \SI{0.1}{\giga\watt} (oil-fired) are decommissioned in our baseline scenario.
    
    Revenues from electricity net exports partly compensate for Austria's system costs. 
    On annual balance, Austria exports \SI{10.6}{\tera\watt\hour} of electricity to Germany. 
    This reversal in the current account\footnote{In 2016, Austria's net imports of electricity amounted to \SI{7.2}{\tera\watt\hour} or \SI{9.9}{\percent} of its total electricity consumption \cite{StatistikAustria2020}.} 
    is a necessary consequence of the government's policy target.
    While (almost) as much electricity is generated from domestic renewable sources as is being consumed, fossil thermal generation units are still required to generate heat and to substitute for renewable generation in times of low electricity generation from intermittent renewable sources.
    
    \begin{figure*}[h!t]
        \centering
        \includegraphics[width=\textwidth]{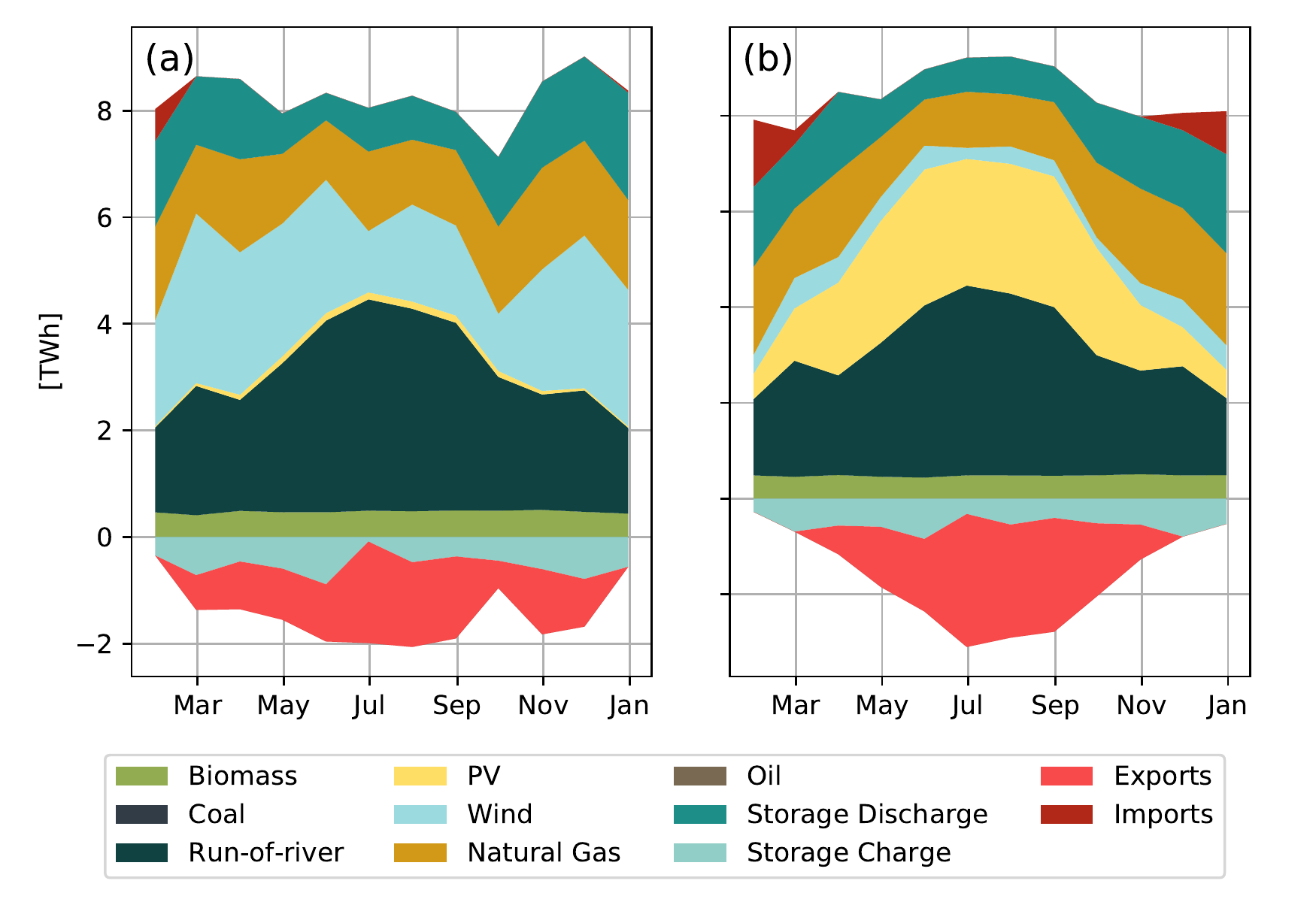}
        \caption{Monthly electricity generation in Austria by source (2030) for (a) maximum wind power and (b) maximum solar power}
        \label{fig:monthly-gen-el}
    \end{figure*}
    
\begin{figure*}[ht]
        \centering
        \includegraphics{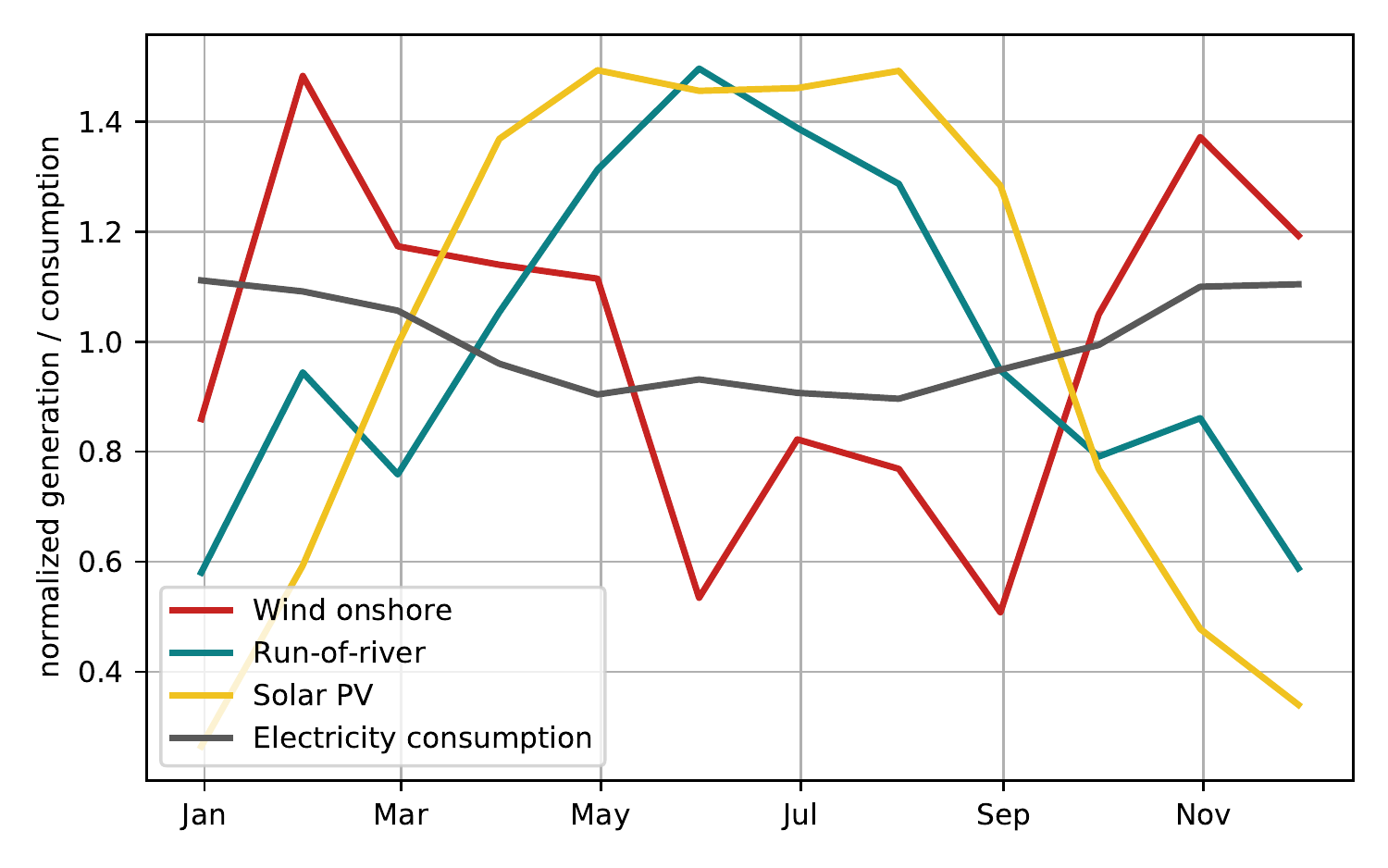}
        \caption{Seasonal pattern of electricity generation from intermittent sources and electricity consumption (monthly average relative to annual average) in Austria in 2016}
        \label{fig:seasonal}
    \end{figure*}

    In the unconstrained baseline scenario with all additional renewable electricity generation coming from wind power, Austria is a net exporter of electricity throughout most of the year. 
    Minor net imports occur in January, as is shown in \figref[a]{monthly-gen-el}. 
    Net exports are relatively stable over the year, as the seasonality of onshore wind power generation is favorably complementary to generation from pre-existing run-of-river hydropower plants (see \autoref{fig:seasonal}), which historically dominate Austria's power generation mix. 
    Austria tends to export electricity to Germany when domestic wind power generation is high (Pearson correlation coefficient $r(X_{AT},wind_{AT})=0.46$, significant at the \SI{1}{\percent} level for hourly data), and to import electricity from Germany when German electricity generation from solar PV is high ($r(I_{AT},PV_{DE})=0.23$, significant at the \SI{1}{\percent} level).
    Overall, Austria's system cost net of the electricity trade balance equals \SI{1.9}[\euro]{} billion, while air pollution costs (excluding the cost of \ce{CO2} emissions) amount to \SI{226}[\euro]{} million.

\subsection{Restricting wind power} \label{subsec:restricting-wind}
\begin{figure*}[ht]
        \centering
        \includegraphics{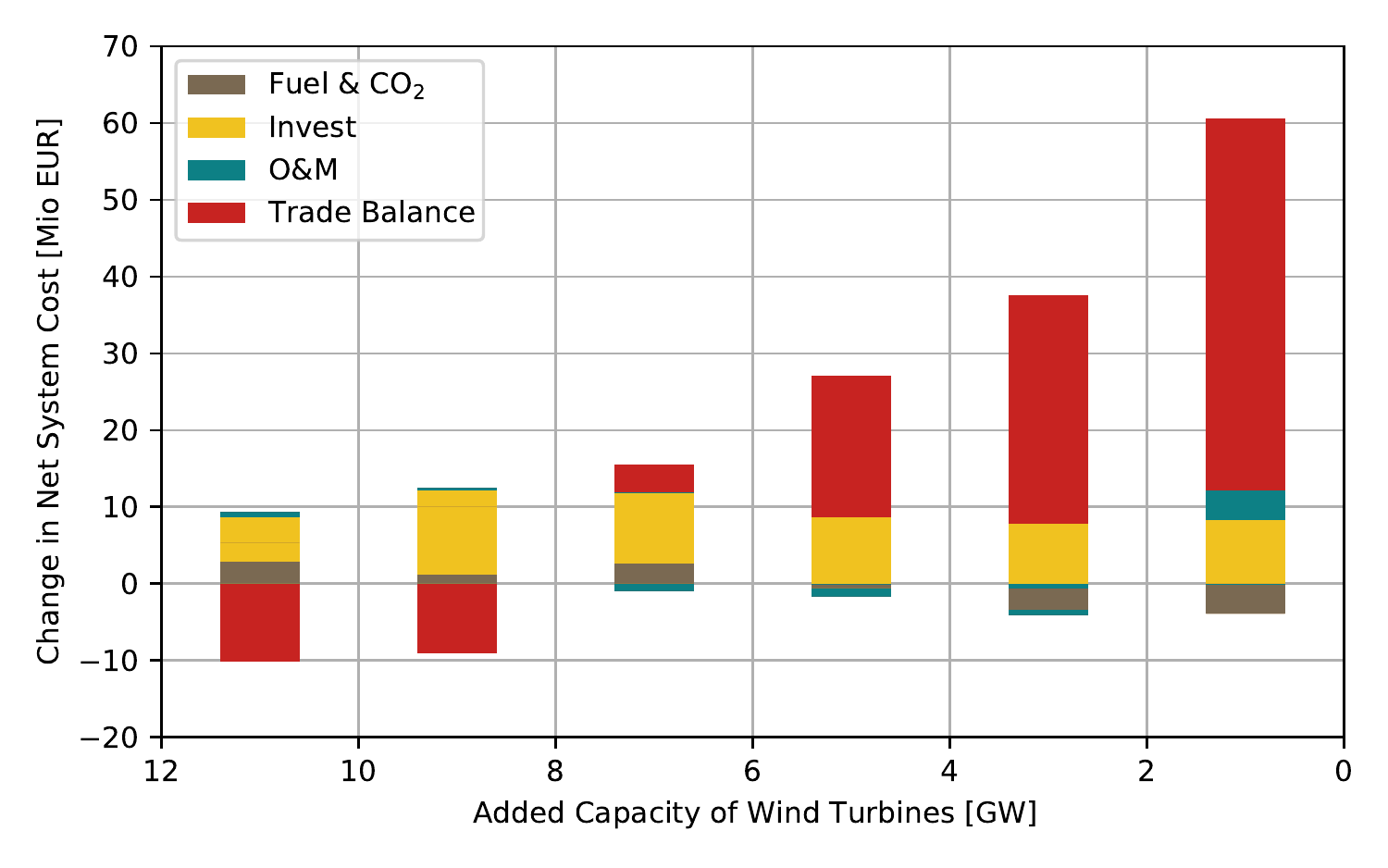}
        \caption{Change in Austria's annual net system cost by components induced by a gradual tightening of the upper limit on wind power installations in the baseline scenario with a \ce{CO2} price of \euro$25$ per tonne. Investment cost includes investment in RET.}
        \label{fig:cost_components}
    \end{figure*}

    With the gradual restriction of wind turbine additions, electricity is increasingly sourced from solar PV. 
    This induces changes in the composition and the operation of the electricity system. 
    As long as the policy constraint of generating a given amount of electricity from renewable sources is binding, any decline in wind power generation must be offset by a corresponding increase in power generation from an alternative renewable power generation technology. 
    Thus, the installed capacity of solar PV increases by \SI{1.98}{\giga\watt} for each \si{\giga\watt} reduction in installed wind power capacity. 
    The corresponding change in investment cost is indicated in \autoref{fig:cost_components}. 
    (The corresponding changes in generation and trade are displayed in \autoref{fig:app_dispatch}.)
    Adding some solar PV to the generation mix reduces the variance of total intermittent electricity generation somewhat, while the impact on minimal intermittent generation is minor.
    In effect, resource adequacy can be achieved with slightly lower thermal capacity in our baseline scenario. 
    More thermal capacity can be decommissioned so that O\&M cost decreases slightly. 
    This reverses as wind power is phased out completely. 
    The variance of intermittent electricity generation is being reduced only marginally, but minimum intermittent generation declines stronger.
    Thus, an electricity system dominated by solar PV requires slightly larger thermal generation capacities. 
    In effect, fewer thermal units can be decommissioned so that system-wide O\&M costs increase.
    
    Changes in the cost of fuel and \ce{CO2} emissions are closely linked to trade flows. 
    As wind power deployment is gradually limited to \SI{6}{\giga\watt} from the initial unconstrained optimum, net exports increase by \SI{400}{\giga\watt\hour} in the baseline scenario. 
    The additional electricity exported is sourced from thermal units (between \SIrange{46}{73}{\percent} of the increase in exports) and lower curtailment of intermittent generation (between \SIrange{55}{28}{\percent}). 
    The increase in thermal generation comes in line with a \SI{94}{\kilo\tonne} increase in \ce{CO2} emissions. 
    A further limitation of wind power below \SI{6}{\giga\watt} leads to a \SI{30}{\percent} (\SI{+260}{\giga\watt\hour}) increase in curtailed electricity and a \SI{350}{\giga\watt\hour} decline in thermal electricity generation.
    These changes come in line with a reduction of electricity net exports to Germany by \SI{810}{\giga\watt\hour}. 
    Moreover, the average price of imports increases by \SI[per-mode=symbol,sticky-per, bracket-unit-denominator=false]{4.69}[\euro]{\per\mega\watt\hour}, while the average price of exports declines by \SI[per-mode=symbol,sticky-per, bracket-unit-denominator=false]{1.76}[\euro]{\per\mega\watt\hour}.
    Taken together, this weakens Austria's electricity trade balance with Germany by \SI{95}[\euro]{} million. 
    The move towards electricity generation from solar PV also leads to a shift in the seasonal pattern of electricity generation, imports, and exports. \autoref{fig:monthly-gen-el} shows the monthly electricity generation and electricity exchange for the two extreme scenarios in which additional adjustable renewable electricity generation is entirely sourced from wind power (\figref[a]{monthly-gen-el}) or solar power (\figref[b]{monthly-gen-el}).
    Comparing these extremes, reduced thermal power generation in the wake of lower exports, particularly in winter months, results in a \SI{162}{\kilo\tonne} decline of \ce{CO2} emissions from the Austrian power sector.
    However, total emissions in Austria and Germany are increasing as wind power is replaced by solar PV. 
    The reduction of Austria's \ce{CO2} emissions is possible only because thermal electricity generation, largely from lignite, expands in Germany. 
    Thus, total system emissions increase by \SI{2.3}{\mega\tonne} or approximately \SI{220}{\kilo\tonne} for each \si{\giga\watt} wind power avoided.
    
\subsection{The opportunity cost of wind power in Austria} \label{subsec:opportunity-cost-wind}
    As laid out in \autoref{subsec:restricting-wind}, solar PV and wind power are not perfectly substitutable in Austria. 
    Thus, diverting from capacities deployed in optimum comes at an opportunity cost. 
    This opportunity cost is relatively small for the whole Austro-German market area. 
    With the complete replacement of wind power by solar PV in Austria, total system cost in Austria and Germany increases by \SI{0.5}{\percent}, while external air pollution cost increases by \SI{3.3}{\percent}. 
    However, the annual Austrian system cost net of trade increase by \SI{6.8}{\percent} or \SI{127}[\euro]{} million, while air pollution cost increases by \SI{75}[\euro]{} million or \SI{33}{\percent}. 
    Relating this cost increase to the change in installed wind power capacity allows us to estimate the marginal opportunity cost of wind power under the implicit assumption that solar PV does not give rise to external cost. 
    Accordingly, we estimate the marginal opportunity cost to increase by \SI[per-mode=symbol,sticky-per, bracket-unit-denominator=false]{2666}[\euro]{\per\mega\watt} wind power foregone averaged across the considered \ce{CO2} price scenarios and assuming PV overnight cost of \SI[per-mode=symbol,sticky-per, bracket-unit-denominator=false]{625}[\euro]{\per\mega\wattpeak}.
    The opportunity cost is small for the first MW of wind power displaced and, depending on the price of CO2, reaches between \SI{32400}[\euro]{} and \SI{41200}[\euro]{} for the last \si{\mega\watt} of wind power displaced.
    
    Given these estimates, we can calculate the valuation of undisturbed landscapes implied by announced government policies. 
    The government intends to source an additional \SI{10}{\tera\watt\hour} annually from wind power, which corresponds to an additional \SI{5.05}{\giga\watt} wind turbines being installed.
    At this level of wind power deployment, the opportunity cost of wind power averaged over all considered \ce{CO2} price scenarios is approximately \SI[per-mode=symbol,sticky-per, bracket-unit-denominator=false]{22500}[\euro]{\per\mega\watt} and year. 
    The present value of the cost of undisturbed landscapes implied by government policies amounts to approximately \SI{1.3}[\euro]{} million over the lifetime of a standard \SI{3.5}{\mega\watt} wind turbine at a \SI{5}{\percent} discount rate. 
    \begin{figure*}[h!t]
        \centering
            \includegraphics{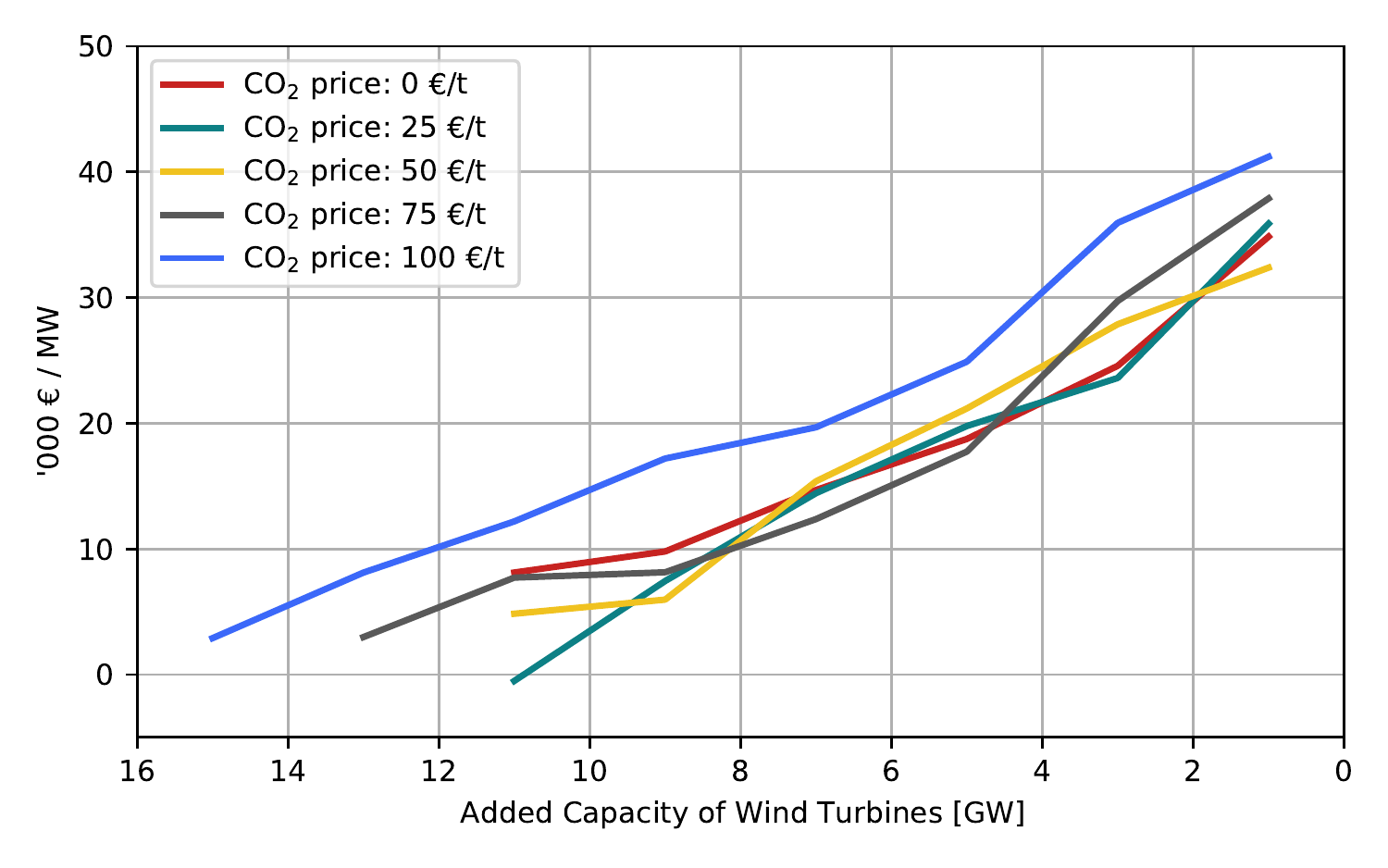}
        \caption{Annual opportunity cost of wind power in Austria, assuming PV overnight cost of \SI[per-mode=symbol,sticky-per, bracket-unit-denominator=false]{625}[\euro]{\per\kilo\wattpeak} (2030)}
        \label{fig:opportunity-cost-base}
    \end{figure*}

\subsection{Sensitivity Analysis} \label{subsec:sensitivity}
    To assess the robustness of our results, we analyze the sensitivity of model outcomes to changes in parameters that can be expected to have a particular influence. The effects of changes in capital cost and electricity transmission capacities are reported below. 
    
    \begin{figure}[h!t]
        \centering
        \includegraphics{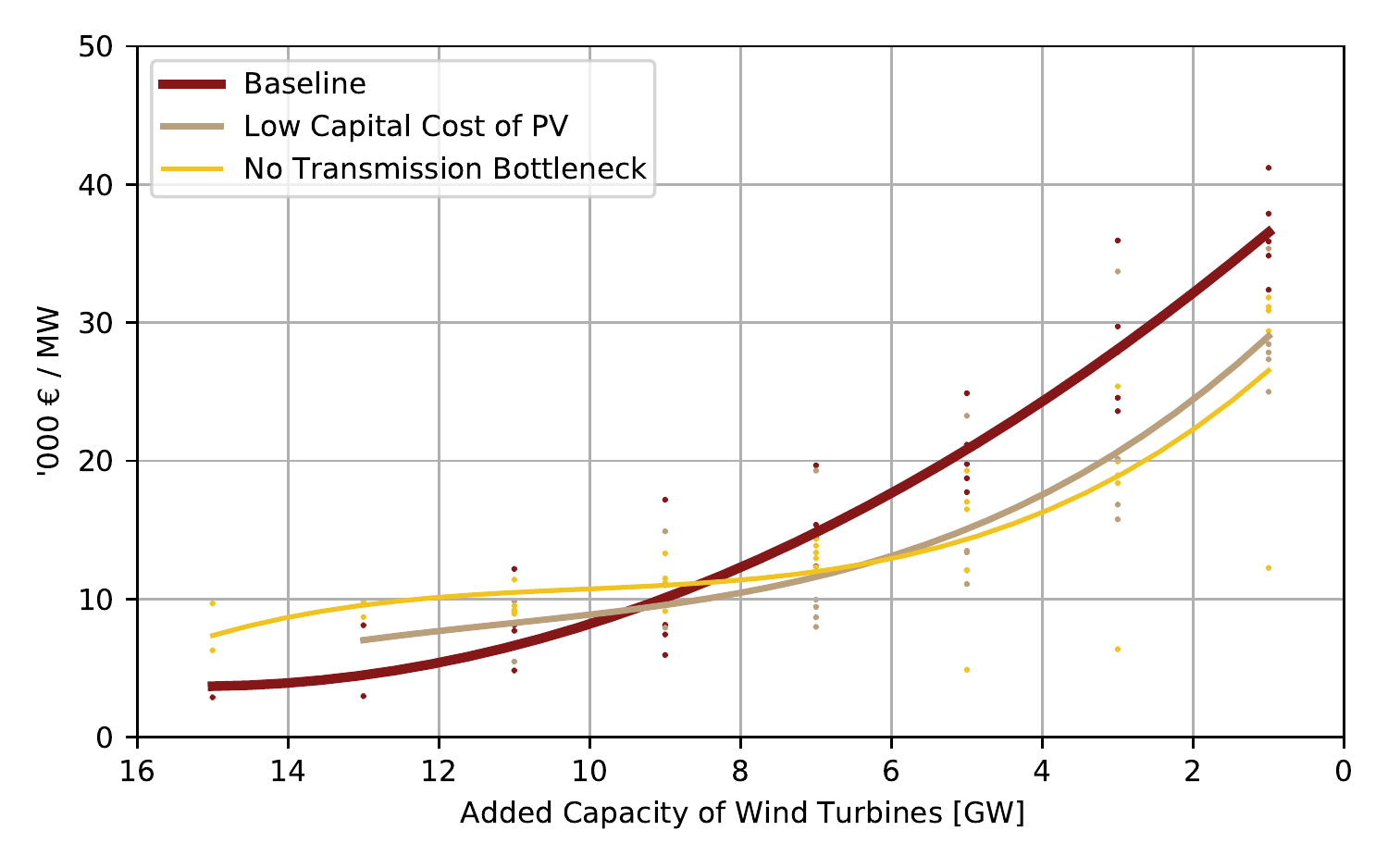}
        \caption{Comparison of the annual opportunity cost of wind power in Austria derived from sensitivity analyses for low capital cost of PV (\SI[per-mode=symbol,sticky-per, bracket-unit-denominator=false]{560}[\euro]{\per\kilo\wattpeak}) and high interconnector capacity (\SI{10}{\giga\watt}) with baseline. Each point represents a specific combination of a \ce{CO2} price assumption and an upper limit on deployable wind power. Lines represent cubic fit to scenario results.}
        \label{fig:sensitivity}
    \end{figure}
    
\subsubsection{Sensitivity to capital cost assumptions}
    To assess the sensitivity of results to changes in the assumed capital cost of wind turbine and solar PV installations, we gradually lower the cost of solar PV from our baseline assumption of \SI[per-mode=symbol,sticky-per, bracket-unit-denominator=false]{625}[\euro]{\per\kilo\wattpeak}, corresponding to a mix of \SI{50}{\percent} rooftop and \SI{50}{\percent} open space installations at 2030 prices, to \SI[per-mode=symbol,sticky-per, bracket-unit-denominator=false]{380}[\euro]{\per\kilo\wattpeak} which corresponds to \SI{100}{\percent} open-space utility-scale PV according to our baseline assumptions. 
    To account for uncertainty in price projections, we lower PV capital cost further down to \SI[per-mode=symbol,sticky-per, bracket-unit-denominator=false]{275}[\euro]{\per\kilo\wattpeak}, which Vartiainen et al.~\cite{Vartiainen2019} project for open-space utility-scale solar PV projects in 2030.
    
    \begin{figure}[h!t]
        \centering
        \includegraphics{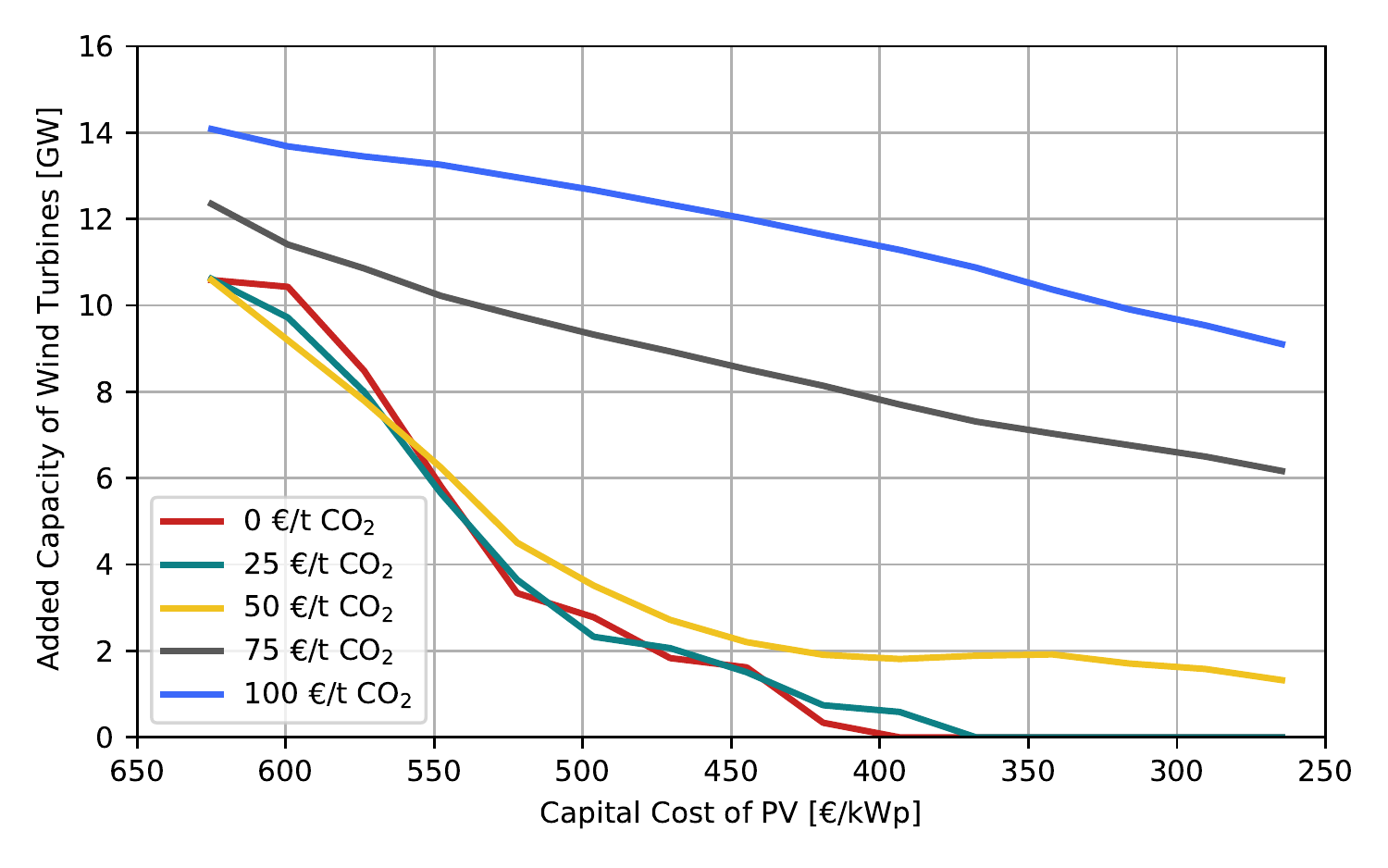}
        \caption{Optimal deployment of wind power in Austria conditional on the capital cost of solar PV}
        \label{fig:opt_wind_cap}
    \end{figure}
    
    As illustrated by \autoref{fig:opt_wind_cap}, we find the optimal level of wind power deployment declining in the capital cost of PV. 
    The marked difference in the effect of the capital cost of solar PV on the optimal level of wind power deployed in Austria is a result of the interactions between a renewables target and a carbon price.
    \ref{subsec:interactions} provides further details about these interactions.
    For emission prices of \SI[per-mode=symbol,sticky-per, bracket-unit-denominator=false]{50}[\euro]{\per\tonne\coo} or lower, this decline is particularly pronounced, with a reduction of the optimally deployed wind power capacity by approximately \SI{0.6}{\giga\watt} for each \SI[per-mode=symbol,sticky-per, bracket-unit-denominator=false]{10}[\euro]{\per\kilo\wattpeak} decline in the capital cost of solar PV. 
    At capital costs of \SI[per-mode=symbol,sticky-per, bracket-unit-denominator=false]{400}[\euro]{\per\kilo\wattpeak} or lower, solar PV covers all additionally required renewable energy generation if the price of \ce{CO2} is at \SI[per-mode=symbol,sticky-per, bracket-unit-denominator=false]{25}[\euro]{\per\tonne} or lower.
    For a carbon price of \SI[per-mode=symbol,sticky-per, bracket-unit-denominator=false]{50}[\euro]{\per\tonne\coo}, about \SI{2}{\giga\watt} of wind power is deployed in optimum at this capital cost. 
    As capital costs fall further, deployed wind power capacity descends slowly, approaching zero as PV capital cost reach \SI[per-mode=symbol,sticky-per, bracket-unit-denominator=false]{275}[\euro]{\per\kilo\wattpeak}.
    With \ce{CO2} price at \SI[per-mode=symbol,sticky-per, bracket-unit-denominator=false]{75}[\euro]{\per\tonne} or higher, when government policy is not binding anymore, deployed wind power capacity remains sturdy. 
    Even at a capital cost of PV as low as \SI[per-mode=symbol,sticky-per, bracket-unit-denominator=false]{275}[\euro]{\per\kilo\wattpeak}, added wind power capacity tops \SI{6}{\giga\watt}. 
    Moreover, it is worth noting that the impact of the lower PV capital cost on \ce{CO2} emissions largely depends on the prevailing \ce{CO2} price level.
    With emission prices of \SI[per-mode=symbol,sticky-per, bracket-unit-denominator=false]{25}[\euro]{\per\tonne\coo} or lower, a halving of PV capital cost reduces total system emissions by only \SI{2}{\percent}.
    In comparison, emission reductions of about \SI{15}{\percent} are realized when \ce{CO2} is priced at least at \SI[per-mode=symbol,sticky-per, bracket-unit-denominator=false]{50}[\euro]{\per\tonne}.
    
    In line with less wind power capacities deployed, the opportunity cost of wind power declines. 
    Given PV's capital cost of \SI[per-mode=symbol,sticky-per, bracket-unit-denominator=false]{560}[\euro]{\per\kilo\wattpeak}, a \SI{10}{\percent} reduction from our baseline, we estimate the opportunity cost of wind power at \SI{5.05}{\giga\watt} added at \SI[per-mode=symbol,sticky-per, bracket-unit-denominator=false]{17500}[\euro]{\per\mega\watt} wind power avoided, a \SI{21.9}{\percent} reduction compared to baseline. 
    However, the reduction in opportunity cost is smaller when restrictions on wind power deployment are tighter. 
    The last megawatt of wind power foregone comes at an opportunity cost of approximately \SI{27200}[\euro]{}, as long as emitting a tonne of \ce{CO2} costs \SI{75}[\euro]{} or less.
    
\subsubsection{Sensitivity to transmission constraints} \label{subsubsec:sens_transmission}
    Following a decision by ACER, electricity transmission capacity between Austria and Germany was lowered from an initially (virtually) unlimited capacity to \SI{4.9}{\giga\watt} in 2018. 
    In consequence, the previously unified Austro-German electricity market zone was split up, allowing for diverging wholesale electricity prices.
    The cap in transmission capacity was implemented to prevent so-called "loop flows" of electricity from Northern Germany through Eastern Europe to Austria and Southern Germany. 
    These loop flows arose from an imbalance of supply in the North and demand in the South, which could all too often not be resolved by the frequently congested inner-German transmission grid. 
    
    \begin{figure}[t]
        \centering
        \includegraphics{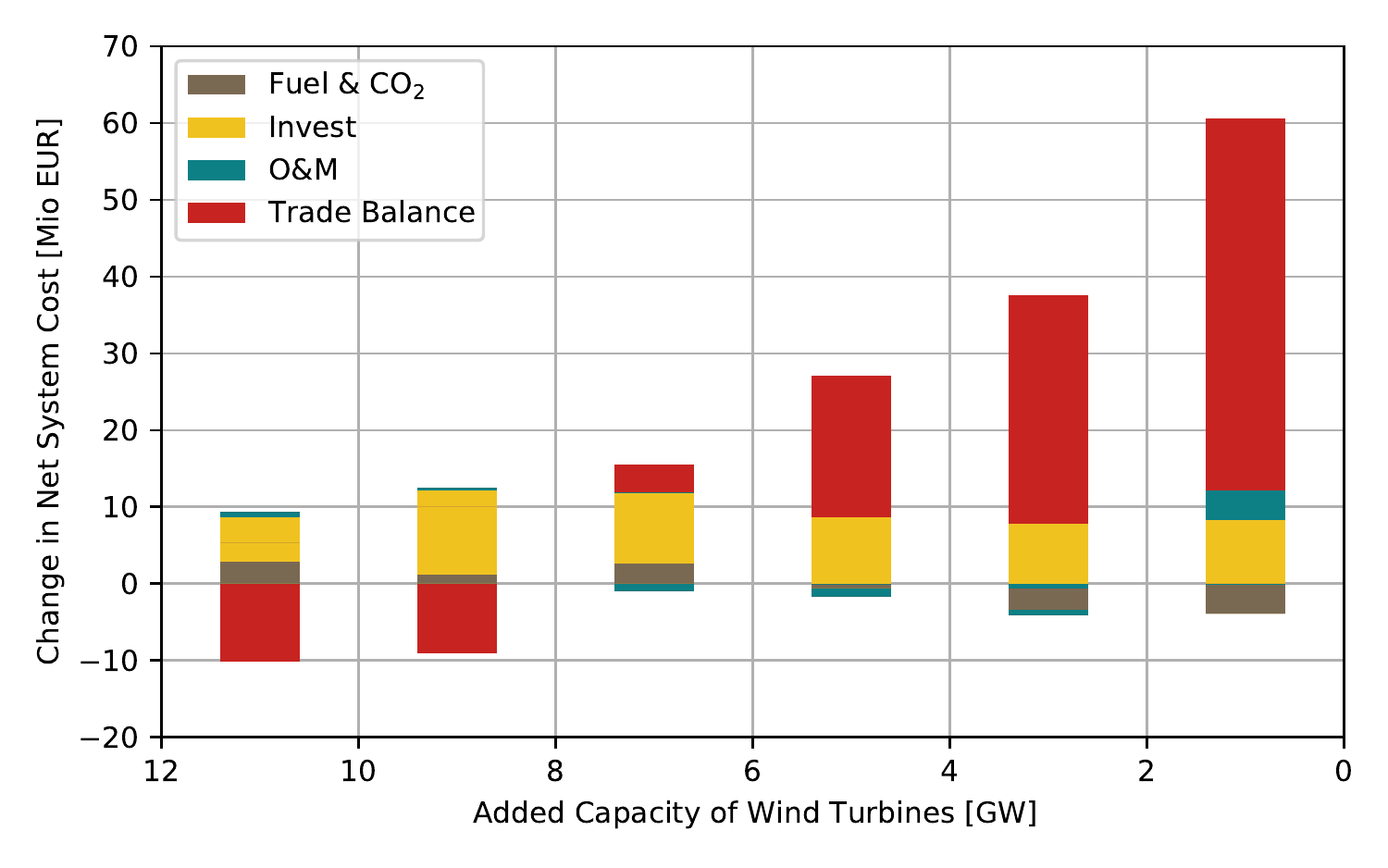}
        \caption{Change in Austria's annual net system cost by components with \SI{10}{\giga\watt} transmission capacity and a \ce{CO2} price of \SI[per-mode=symbol,sticky-per, bracket-unit-denominator=false]{25}[\euro]{\per\tonne}}
        \label{fig:sens_interconn_costco}
    \end{figure}

    Given that current energy policy goals turn Austria from a net importer into a net exporter of electricity, we have analyzed the effects of a potential lifting of ACER's decision to cap electricity transmission by setting transmission capacity to \SI{10}{\giga\watt}.
    Under this assumption, the opportunity cost of wind power is higher compared to baseline at low penetrations of solar PV.
    However, the increase in opportunity cost as the PV penetration increases is less pronounced compared to baseline.
 
    Higher transmission capacity allows (fossil) power plants to generate and export electricity at times when interconnectors with lower transmission capacity would be congested. 
    In consequence, use of fossil fuels for thermal power generation and \ce{CO2} emissions are higher (by \SIrange{390}{670}{\kilo\tonne} at a carbon price of \SI[per-mode=symbol,sticky-per, bracket-unit-denominator=false]{25}[\euro]{\per\tonne\coo}) compared to our baseline with lower interconnection capacity.
    Moreover, thermal capacity is not being shut down (as in the baseline scenario), so that changes in O\&M cost are very small. 

    Higher interconnector capacity allows for higher quantities of imports and exports. 
    Indeed, Austrian imports are up by \SI[per-mode=symbol,sticky-per, bracket-unit-denominator=false]{3.1}{\tera\watt\hour\per\year}, while exports increase by about \SI[per-mode=symbol,sticky-per, bracket-unit-denominator=false]{2.6}{\tera\watt\hour\per\year}, compared to the unconstrained baseline scenario.

    Even though net exports decline, Austria's (net) export surplus increases, mostly on the back of a lower average import price for electricity from Germany.
    Moreover, increased interconnector capacity leads to a trade balance that is less responsive to changes in installed RET capacities. 
    Thus, the opportunity cost of wind power are higher at at high wind power penetration, but lower at low wind power penetration (see \autoref{fig:sens_interconn_costco}).

\subsection{Limitations of the Analysis} \label{subsec:limitations}
    It is worth noting that our analysis is not spatially resolved, which has important consequences.
    Our analysis of the opportunity cost of wind power does not depend on a specific spatial allocation of RET or a specific grid topology. Hence, our results are generalizable to the extent that any assessment of specific RET expansion plans must consider the effects of spatially distributed RET on resource quality and grid cost.
    However, there are good reasons to believe these factors will not change the results of our analysis substantially.
    First, we implicitly assume that wind and solar resource quality at locations developed in 2016 is representative of resource quality at locations developed by 2030, i.e. after generation increased by \SI{21}{\tera\watt\hour}.
    However, Hoeltinger et al. \cite{Hoeltinger2016}, for example, identified technically feasible wind resource potential of \SI{45}{\tera\watt\hour} in Austria which is of better or similar quality than we assumed in our analysis. 
    Hence, \SI{21}{\tera\watt\hour} additional electricity generation from wind power is feasible under our assumptions.
    Yet, technical potentials are currently restricted by law in four of Austria's nine federal states. These regulations limit total wind energy potential in these states to approximately \SI{11}{\tera\watt\hour}, of which around \SI{6}{\tera\watt\hour} are already used.
    Thus, it will be necessary to lift these restrictions and amend current regulations even under current policy objectives. 
    
    Second, we implicitly assume that the cost of grid expansion does not differ between scenarios relying entirely on wind power and scenarios relying on solar PV exclusively.
    Here, it makes a difference whether solar PV is realized as rooftop systems in distribution grids or as utility-scale open-space systems that are connected to transmission grids. 
    In the latter case, there is little reason to believe grid cost would differ substantially from wind power expansion, which also happens in transmission grids.
    \cite{Horowitz2018} survey the nascent literature on the cost of integrating distributed solar PV systems in distribution grids. 
    These costs are found to depend on numerous and uncertain factors, leading the authors to the conclusion that "no generalized `cost of grid integration' for PV can be obtained".
    Nevertheless, \cite{Fuerstenwerth2015} estimate grid integration cost of solar PV in Europe at \SI[per-mode=symbol,sticky-per, bracket-unit-denominator=false]{6}[\euro]{\per\mega\watt\hour} in distribution grids, and at \SI[per-mode=symbol,sticky-per, bracket-unit-denominator=false]{1.5}[\euro]{\per\mega\watt\hour} in transmission grids. 
    Estimates for specific grids, however, vary widely.
    Given the available evidence, we are inclined to assume that any grid cost advantage of PV over onshore wind is likely small.
    
    Lastly, in our analysis we focus on the time-horizon up to 2030.
    Thus, our findings are not necessarily transferable to the (almost) fully decarbonized electricity systems we might establish in the more distant future. 
    Yet, fundamental properties of renewable resources, such as seasonal and diurnal patterns in wind speeds or solar irradiation, will persist over these time frames. 
    Thus, we expect to find positive opportunity costs of wind power also in a decarbonized Austrian electricity system, particularly as future options to facilitate the system integration of variable renewables that we did not model, are unlikely to come at significantly lower cost than the included options.
 
\section{Conclusions and Policy Implication} \label{sec:conclusions-policy-implication}
    Not disturbing landscapes can come at considerable opportunity cost, as our analysis reveals.
    How high this cost is, depends on the valuation of \ce{CO2} and the cost of the best alternative to wind power.
    If the cost of \ce{CO2} is low and the best alternative solar PV is realized at utility-scale in open space, wind power's system cost advantage vanishes in Austria.
    If, on the other hand, the cost of \ce{CO2} is high, if open-space PV itself causes negative external effects at the local level, or if there is a preference for rooftop solar PV, the opportunity cost of wind power is worthy of consideration.
    
    Given the current policy target of adding capacities for generating \SI{10}{\tera\watt\hour} wind energy, the present value of the cost of undisturbed landscapes amounts to at least \SI{995000}[\euro]{} over the lifetime of each \SI{3.5}{\mega\watt} wind turbine not erected.\footnote{We assume a lifetime of 30 years and a discount factor of \SI{5}{\percent}.}

    While these numbers indicate that in the absence of local wind turbine externalities, the government targets for RET deployment aren't economically efficient, a complete policy evaluation must weigh the cost of undisturbed landscapes against its benefits arising from the preservation of current landscapes aesthetics, for example.
    Yet, any significant expansion of renewable electricity generation technologies will interfere with landscapes.
    Relying completely on open-space solar PV would require about \SI{315}{\square\kilo\meter} land\footnote{We assume land usage of \SI{15}{\square\meter} per \si{\kilo\wattpeak} of non-tracking silicon wafer technology.} 
    for replacing the approximately 3\,030 wind turbines of the \SI{3.5}{\mega\watt} class that would be needed to meet policy goals by wind power alone.
    
    For a comprehensive policy evaluation, our findings need to be complemented by a spatially highly resolved analysis of the local cost of disturbed landscapes.
    Together, such findings would allow determining the socially optimal deployment of renewable energy generation technologies. 
    Here, we see a fruitful field for further research. 
    
    Attention should also be paid to the distributive consequences of wind power expansion in Austria.
    As we have shown, a large-scale expansion of wind turbines comes with an increase in electricity exports, i.e. at high penetration some wind power capacity is mostly used to generate electricity for export.
    Revenues from these export are accruing to the owners of power generation assets, while negative external effects of wind turbines have to be borne by local residents.
    This imbalance could be narrowed down by policies requiring wind turbine owners to share wind turbine income with affected residents, either directly or through transfers.
    Fair sharing of benefits and burdens has the potential to foster acceptance of wind power \citep{Scherhaufer2017}.
  
    \section*{Acknowledgements}
    Sebastian Wehrle would like to thank Katharina Gruber, Peter Regner and Claude Klöckl for helpful comments on early drafts of the manuscript.
    We gratefully acknowledge support from the European Research Council (“reFUEL” ERC-2017-STG 758149).
    
    \newpage
    \bibliography{asparagus}
    \bibliographystyle{elsarticle-harv}
    
    \newpage
    \appendix
\section{Change in dispatch and trade for a gradual tightening of the upper limit on wind power installations}
\begin{figure}[h!]
    \centering
    \begin{subfigure}[b]{0.49\textwidth}
        \centering
        \includegraphics[width=\textwidth]{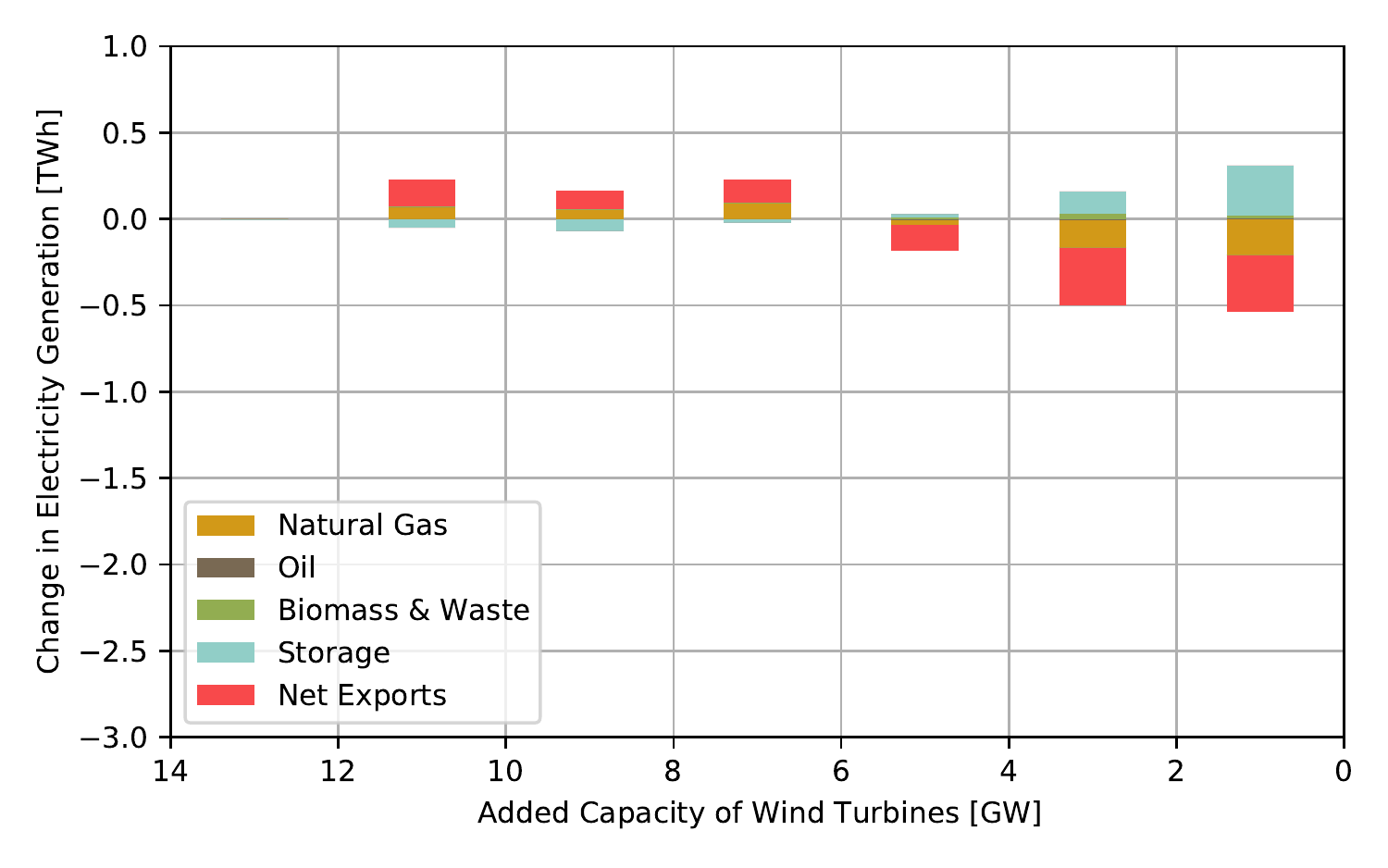}
        \caption{Carbon price of \SI[per-mode=symbol,sticky-per, bracket-unit-denominator=false]{25}[\euro]{\per\tonne\coo}}
    \label{fig:app_dispatch_a}
    \end{subfigure}
    \hfill
    \begin{subfigure}[b]{0.49\textwidth}
        \centering
        \includegraphics[width=\textwidth]{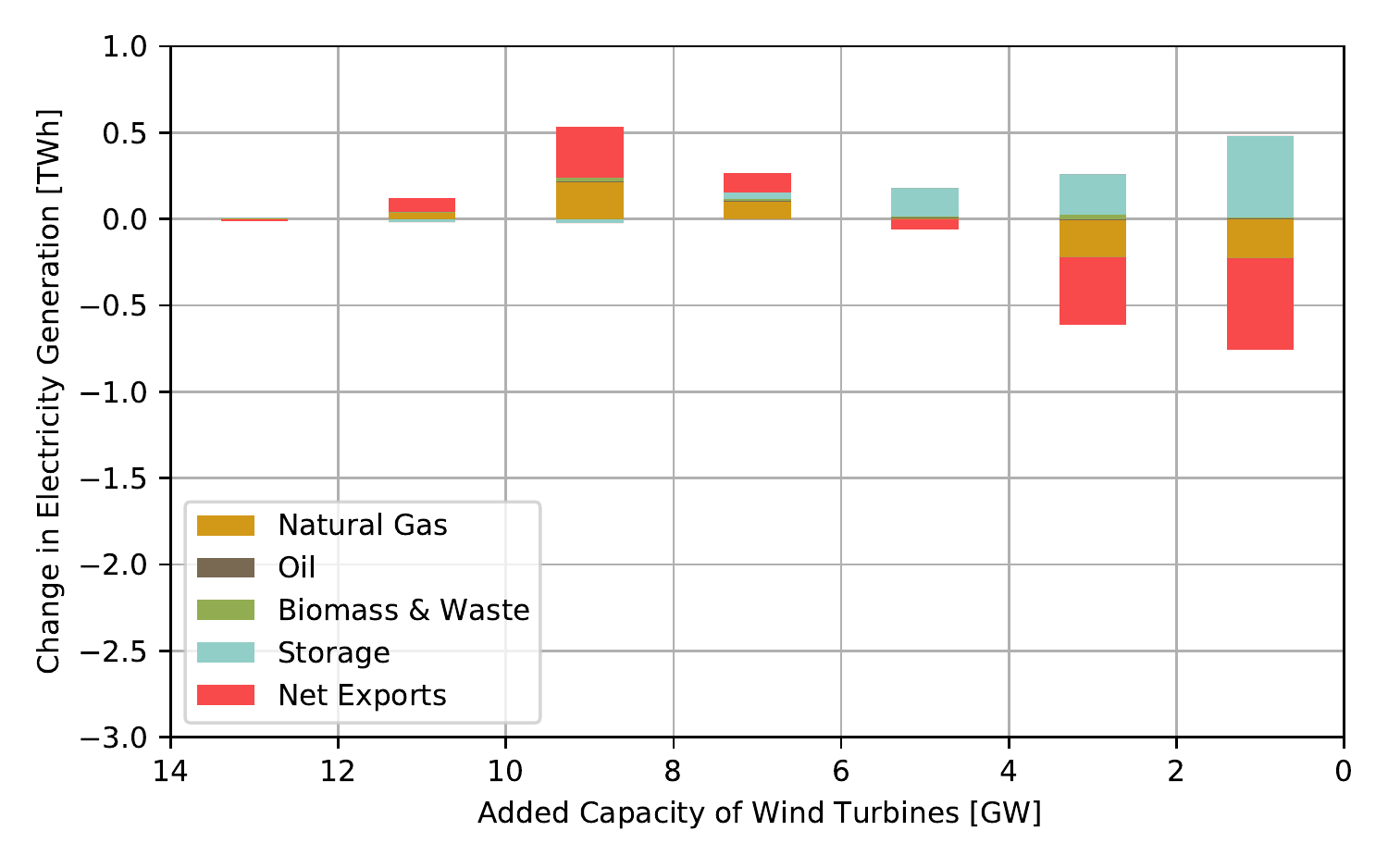}
        \caption{Carbon price of \SI[per-mode=symbol,sticky-per, bracket-unit-denominator=false]{50}[\euro]{\per\tonne\coo}}
    \label{fig:app_dispatch_b}
    \end{subfigure}
    \hfill
    \begin{subfigure}[b]{0.49\textwidth}
        \centering
        \includegraphics[width=\textwidth]{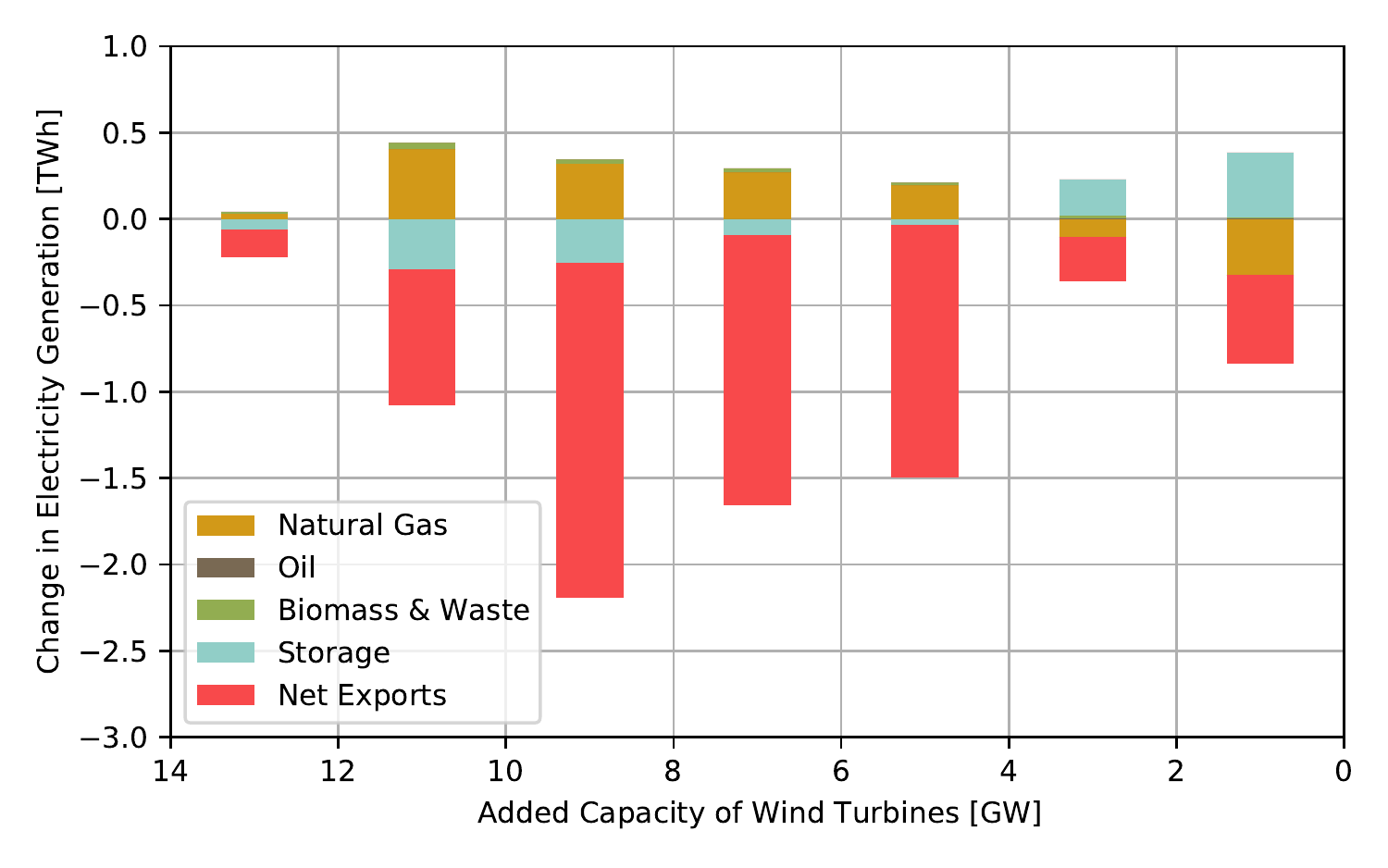}
        \caption{Carbon price of \SI[per-mode=symbol,sticky-per, bracket-unit-denominator=false]{75}[\euro]{\per\tonne\coo}}
    \label{fig:app_dispatch_c}
    \end{subfigure}
    \hfill
    \begin{subfigure}[b]{0.49\textwidth}
        \centering
        \includegraphics[width=\textwidth]{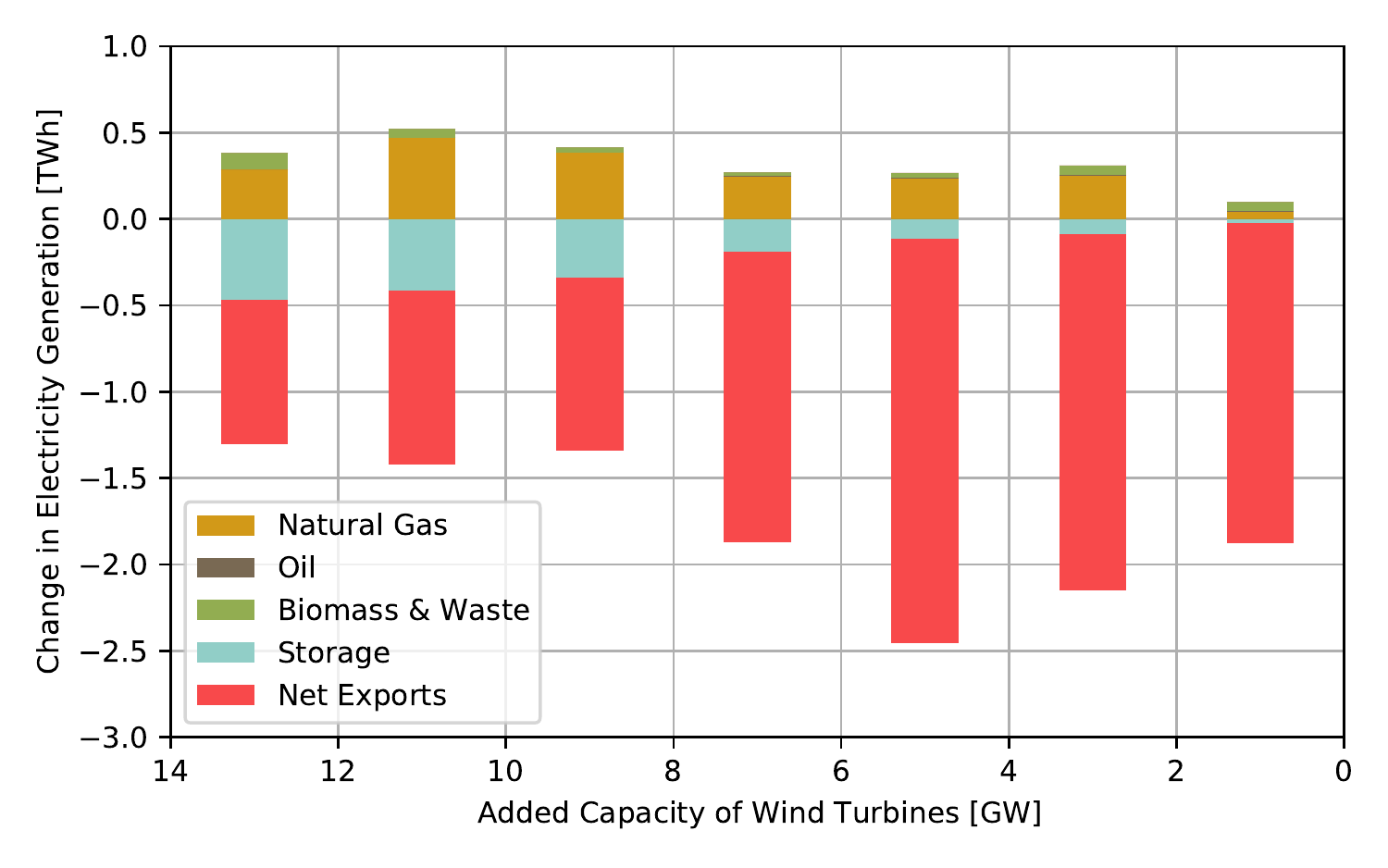}
        \caption{Carbon price of \SI[per-mode=symbol,sticky-per, bracket-unit-denominator=false]{100}[\euro]{\per\tonne\coo}}
    \label{fig:app_dispatch_c}
    \end{subfigure}

    \caption{Change in Austria's annual electricity generation by fuel and electricity exchange with Germany induced by a gradual tightening of the upper limit on wind power installations.}
    \label{fig:app_dispatch}
\end{figure}

\section{The interaction of the policy target with \ce{CO2} prices} \label{subsec:interactions}
    The Austrian government seeks to achieve its renewable electricity generation target through the provision of technology-specific subsidies. 
    The subsidization of renewable electricity generation induces an expansion of installed RET capacities. 
    Pricing carbon dioxide emissions would be an alternative policy to incentivize investment in renewable energy generation, as carbon pricing would improve the competitiveness of carbon-free generators versus their fossil counterparts.
    The higher the price of \ce{CO2}, the stronger the incentive for investment in renewable energy technologies. 
    Hence, there is a \ce{CO2} price that is equivalent to reaching the same emission reduction. 
    In our baseline scenario, the policy reduces \ce{CO2} emissions by \SI{1.2}{\mega\tonne} versus the same scenario without the policy constraint. 
    An emission reduction of similar size could also be reached through raising the \ce{CO2} price to approximately \SI[per-mode=symbol,sticky-per, bracket-unit-denominator=false]{40}[\euro]{\per\tonne}. 
    However, the resulting system is not equivalent in terms of installed capacities of electricity generated from domestic renewable sources. 
    Reaching this policy objective would require a carbon price of approximately \SI[per-mode=symbol,sticky-per, bracket-unit-denominator=false]{63}[\euro]{\per\tonne\coo}. 
    For emission prices above this level, the government policy is not binding anymore.
    A major difference between a binding government policy objective and a hypothetical \ce{CO2} price lies in the measures' effect on electricity trade. 
    Under the renewables target, Austria necessarily turns into a net exporter of electricity and benefits from the associated revenues. 
    Under \ce{CO2} pricing, Austria would remain a net importer of electricity and maintain a negative balance of payments in the electricity sector up to a \ce{CO2} price of approximately \SI[per-mode=symbol,sticky-per, bracket-unit-denominator=false]{45}[\euro]{\per\tonne}.

\section{Description of the power system model \emph{medea}} \label{sec:medea-desc}

\subsection{Sets} \label{sets}
    Sets are denoted by upper-case latin letters, while set elements are denoted by lower-case latin letters.

    \begin{longtable}{p{0.15\textwidth}p{0.15\textwidth}p{0.2\textwidth}p{0.45\textwidth}}
        \caption{Sets}\\
        \toprule
        mathematical symbol & programming symbol & description & elements\\
        \midrule
        \endhead
        \bottomrule
        \multicolumn{4}{c}{\textit{continued on next page}} \endfoot
        \bottomrule
        \endlastfoot
        $f \in F$            &\texttt{f}         & fuels & \texttt{nuclear, lignite, coal, gas, oil, biomass, power} \\
        $i \in I$            &\texttt{i}         & energy conversion technologies & \texttt{nuc, lig\_stm, lig\_stm\_chp, lig\_boa, lig\_boa\_chp, coal\_sub, coal\_sub\_chp, coal\_sc, coal\_sc\_chp, coal\_usc, coal\_usc\_chp, coal\_igcc, ng\_stm, ng\_stm\_chp, ng\_ctb\_lo, ng\_ctb\_lo\_chp, ng\_ctb\_hi, ng\_ctb\_hi\_chp, ng\_cc\_lo, ng\_cc\_lo\_chp, ng\_cc\_hi, ng\_cc\_hi\_chp, ng\_mtr, ng\_mtr\_chp, ng\_boiler\_chp, oil\_stm, oil\_stm\_chp, oil\_ctb, oil\_ctb\_chp, oil\_cc, oil\_cc\_chp, bio, bio\_chp, heatpump\_pth}                                 \\
        $h \in H \subset I$    &\texttt{h(i)}    & power to heat technologies & \texttt{heatpump\_pth} \\
        $j \in J \subset I$    &\texttt{j(i)}    & CHP technologies & \texttt{lig\_stm\_chp, lig\_boa\_chp, coal\_sub\_chp, coal\_sc\_chp, coal\_usc\_chp, ng\_stm\_chp, ng\_ctb\_lo\_chp, ng\_ctb\_hi\_chp, ng\_cc\_lo\_chp, ng\_cc\_hi\_chp, ng\_mtr\_chp, ng\_boiler\_chp, oil\_stm\_chp, oil\_ctb\_chp, oil\_cc\_chp, bio\_chp} \\
        $k \in K$               &\texttt{k}       & storage technologies & \texttt{res\_day, res\_week, res\_season, psp\_day, psp\_week, psp\_season, battery} \\
        $l \in L$              &\texttt{l}       & feasible operation region limits & \texttt{l1, l2, l3, l4}\\
        $m \in M$              &\texttt{m}       & energy products & \texttt{el, ht} \\
        $n \in N$               &\texttt{n}       & intermittent generators & \texttt{wind\_on, wind\_off, pv, ror} \\
        $t \in T$              &\texttt{t}       & time periods (hours)               & \texttt{t1, t2, \ldots, t8760}\\
        $z \in Z$              &\texttt{z}       & market zones & \texttt{AT, DE} \\
    \end{longtable}

\newpage
\subsection{Parameters} \label{parameters}
    Parameters are denoted either by lower-case greek letters or by upper-case latin letters.

    \begin{longtable}{p{0.15\textwidth}p{0.3\textwidth}p{0.35\textwidth}p{0.15\textwidth}}
    \caption{Parameters}\\
    \toprule
    mathematical symbol & programming symbol & description & unit\\
    \midrule
    \endhead
    \bottomrule
    \multicolumn{4}{c}{\textit{continued on next page}} \endfoot
    \bottomrule
    \endlastfoot
    $\Bar{\alpha}_{f}$ & \texttt{AIR\_POL\_COST\_FIX(f)} & external cost of air pollution & \SI{}[\euro]{\per\mega\watt}\\
    $\alpha_{f}$ & \texttt{AIR\_POL\_COST\_VAR(f)} & external cost of air pollution & \SI{}[\euro]{\per\mega\watt\hour}\\
        $\delta_{z,zz}$              &\texttt{DISTANCE(z,zz)}                           & distance between countries' center of gravity & km                        \\
        $\varepsilon_{f}$            &\texttt{CO2\_INTENSITY(f)}                        & fuel emission intensity & $\text{t}_{\ce{CO2}}$/ MWh  \\
        $\eta_{i,m,f}$               &\texttt{EFFICIENCY\_G(i,m,f)}                     & power plant efficiency & MWh / MWh                 \\
        $\eta^{out}_{z,k}$           &\makecell[l]{\texttt{EFFICIENCY\_S\_OUT(k)}}      & discharging efficiency &                           \\
        $\eta^{in}_{z,k}$            &\makecell[l]{\texttt{EFFICIENCY\_S\_IN(k)}}       & charging efficiency &                           \\
        $\lambda_{z}$                &\texttt{LAMBDA(z)}                                & scaling factor for peak load &                           \\
        $\mu_{z}$                    &\texttt{VALUE\_NSE(z)}                            & value of lost load & \EUR/ MWh               \\
        $\rho_{z,t,k}$               &\makecell[l]{\texttt{INFLOWS(z,t,k)}}             & inflows to storage reservoirs & MW                        \\
        $\sigma_{z}$                 &\texttt{SIGMA(z)}                                 & scaling factor for peak intermittent generation &                           \\
        $\phi_{z,t,n}$               &\texttt{GEN\_PROFILE(z,t,n)}                      & intermittent generation profile & $[0,1]$                  \\
        $\widehat{\phi}_{z,n}$       &\texttt{PEAK\_PROFILE(z,n)}                       & peak intermittent generation profile & $[0,1]$                   \\
        $\chi_{i,l,f}$               &\texttt{FEASIBLE\_INPUT(i,l,f)}                   & inputs of feasible operating region & $[0,1]$                  \\
        $\psi_{i,l,m}$               &\texttt{FEASIBLE\_OUTPUT(i,l,m)}                  & output tuples of feasible operating region & $[0,1]$                  \\
        $C^{r}_{z,n}$                &\texttt{CAPITALCOST\_R(z,n)}                      & capital cost of intermittent generators (specific, annuity)     & \EUR/ MW                \\
        $C^{g}_{z,i}$                &\texttt{CAPITALCOST\_G(z,i)}                      & capital cost of thermal generators (specific, annuity)          & \EUR/ MW                 \\
        $C^{s}_{z,k}$                &\makecell[l]{\texttt{CAPITALCOST\_S(z,k)}}        & capital cost of storages - power (specific, annuity)            & \EUR/ MW                \\
        $C^{v}_{z,k}$                &\makecell[l]{\texttt{CAPITALCOST\_V(z,k)}}        & capital cost of storages - energy (specific, annuity)           & \EUR/ MW                \\
        $C^{x}$                      &\texttt{CAPITALCOST\_X}                           & capital cost of transmission capacity (specific, annuity)       & \EUR/ MW                \\
        $D_{z,t,m}$                  &\texttt{DEMAND(z,t,m)}                            & energy demand & GW                        \\
        $\widehat{D}_{z,m}$           &\texttt{PEAK\_LOAD(z,m)}                          & peak demand & GW                        \\
        $\widetilde{G}_{z,i}$        &\texttt{INITIAL\_CAP\_G(z,tec)}                   & initial capacity of dispatchable generators & GW                        \\
        $O^{g}_{i}$                  &\texttt{OM\_COST\_G\_VAR(i)}                      & variable O\&M cost of dispatchable generators & \EUR/ MWh               \\
        $O^{r}_{z,n}$                &\texttt{OM\_COST\_R\_VAR(z,n)}                    & variable O\&M cost of intermittent generators & \EUR/ MWh               \\
        $\Bar{O}^{g}_{i}$                  &\texttt{OM\_COST\_G\_QFIX(i)}                     & quasi-fixed O\&M cost of dispatchable generators & \EUR/ MW                \\
        $\Bar{O}^{r}_{z,n}$                &\texttt{OM\_COST\_R\_QFIX(z,n)}                   & quasi-fixed O\&M cost of intermittent generators & \EUR/ MW                \\
        $P^{e}_{t,z}$                &\texttt{PRICE\_CO2(t,z)}                          & \ce{CO2} price &\EUR/$\text{t}_{\ce{CO2}}$ \\
        $P_{t,z,f}$                  &\texttt{PRICE\_FUEL(t,z,f)}                       & fuel price & \EUR/ MWh               \\
        $\widetilde{R}_{z,n}$        &\texttt{INITIAL\_CAP\_R(z,n)}                     & initial capacity of intermittent generators & GW                        \\
        $\widetilde{S}^{out}_{z,k}$  &\makecell[l]{\texttt{INITIAL\_CAP\_S\_OUT(z,k)}}  & initial discharging capacity of storages & GW                        \\
        $\widetilde{S}^{in}_{z,k}$   &\makecell[l]{\texttt{INITIAL\_CAP\_S\_IN(z,k)}}   & initial charging capacity of storages & GW                        \\
        $\widetilde{V}_{z,k}$        &\makecell[l]{\texttt{INITIAL\_CAP\_V(z,k)}}       & initial energy storage capacity &                           \\
        $\widetilde{X}_{z,zz}$       &\texttt{INITIAL\_CAP\_X(z,zz)}                    & initial transmission capacity & GW                        \\
    \end{longtable}

\newpage
\subsection{Variables} \label{variables}
    Variables are denoted by lower-case latin letters.
    
     \begin{longtable}{p{0.15\textwidth}p{0.3\textwidth}p{0.35\textwidth}p{0.15\textwidth}}
    \caption{Variables}\\
    \toprule
    mathematical symbol & programming symbol & description & unit\\
    \midrule
    \endhead
    \bottomrule
    \multicolumn{4}{c}{\textit{continued on next page}} \endfoot
    \bottomrule
    \endlastfoot
        $b_{z,t,i,f}$                &\texttt{b(z,t,i,f)}           & fuel burn for energy generation & \si{\giga\watt} \\
        $c$                          &\texttt{cost\_system}         & total system cost & k\EUR     \\
        $c_{z}$                       &\texttt{cost\_zonal(z)}       & zonal system cost & k\EUR     \\
        $c^{a}_{z,f}$ & \texttt{cost\_air\_pol(z,f)} & total external cost of air pollution & k\EUR  \\
        $c^{b}_{z,t,i}$              &\texttt{cost\_fuel(z,t,i)}    & fuel cost & k\EUR     \\
        $c^{e}_{z,t,i}$              &\texttt{cost\_co2(z,t,i)}     & emission cost & k\EUR     \\
        $c^{om}_{z,i}$               &\texttt{cost\_om\_g(z,i)}     & total o\&m cost of dispatchable generators & k\EUR     \\
        $c^{om}_{z,n}$               &\texttt{cost\_om\_r(z,n)}     & total o\&m cost of intermittent generators & k\EUR     \\
        $c^{g}_{z}$                  &\texttt{cost\_invest\_g(z)}   & capital cost of generators & k\EUR     \\
        $c^{q}_{z}$                  &\texttt{cost\_nse(z)}         & total cost of non-served load & k\EUR     \\
        $c^{r}_{z}$                  &\texttt{cost\_invest\_r(z)}   & capital cost of intermittent generators & k\EUR     \\
        $c^{s,v}_{z}$                &\texttt{cost\_invest\_sv(z)}  & capital cost of storages & k\EUR     \\
        $c^{x}_{z}$                  &\texttt{cost\_invest\_x(z)}   & capital cost of interconnectors & k\EUR     \\
        $e_{z}$                      &\texttt{emission\_co2(z)}     & \ce{CO2} emissions & t\ce{CO2}    \\
        $\widetilde{g}^{+}_{z,i}$    &\texttt{add\_g(z,i)}          & added capacity of dispatchables & GW        \\
        $\widetilde{g}^{-}_{z,i}$    &\texttt{deco\_g(z,i)}         & decommissioned capacity of dispatchables & GW        \\
        $g_{z,t,i,m,f}$              &\texttt{g(z,t,i,m,f)}         & energy generated by conventionals & GW        \\
        $q^{+}_{z,t}$                &\texttt{q\_curtail(z,t)}      & curtailed energy & GW        \\
        $q^{-}_{z,t,m}$               &\texttt{q\_nse(z,t,m)}        & non-served energy & GW        \\
        $\widetilde{r}^{+}_{z,n}$    &\texttt{add\_r(z,n)}          & added capacity of intermittents & GW        \\
        $\widetilde{r}^{-}_{z,n}$    &\texttt{deco\_r(z,n)}         & decommissioned capacity of intermittents & GW        \\
        $r_{z,t,n}$                  &\texttt{r(z,t,n)}             & electricity generated by intermittents & GW        \\
        $\widetilde{s}^{+}_{z,k}$    &\texttt{add\_s(z,k)}          & added storage capacity (power)              & GW        \\
        $s^{in}_{z,t,k}$             &\texttt{s\_in(z,t,k)}         & energy stored in & GW        \\
        $s^{out}_{z,t,k}$            &\texttt{s\_out(z,t,k)}        & energy stored out & GW        \\
        $\widetilde{v}^{+}_{z,k}$    &\texttt{add\_v(z,k)}          & added storage capacity (energy)             & GWh       \\
        $v_{z,t,k}$                  &\texttt{v(z,t,k)}             & storage energy content & GWh       \\
        $w_{z,t,i,l,f}$              &\texttt{w(z,t,i,l,f)}         & operating region weight &           \\
        $\widetilde{x}^{+}_{z,zz}$   &\texttt{add\_x(z,zz)}         & added transmission capacity & GW        \\
        $x_{z,zz,t}$                 &\texttt{x(z,zz,t)}            & electricity net export & GW        \\
    \end{longtable}

\newpage
\subsection{Naming system}
    \begin{table*}[h!]
        \centering
        \begin{threeparttable}
            \caption{Naming System}
            \begin{tabulary}{\textwidth}{LLLLLL}
                \toprule
                & initial capacity\tnote{$\dagger$}    & added capacity\tnote{$\ddagger$}  & decommissioned capacity\tnote{$\ddagger$} & specific investment cost\tnote{$\dagger$} & dispatch\tnote{$\ddagger$}   \\
                \midrule
                thermal units &$\widetilde{G}_{z,i}$                &$\widetilde{g}^{+}_{z,i}$         &$\widetilde{g}^{-}_{z,i}$                &$C^{g}_{z,i}$                            &$g_{z,t,i,m,f}$              \\
                intermittent units &$\widetilde{R}_{z,n}$                &$\widetilde{r}^{+}_{z,n}$        &$\widetilde{r}^{-}_{z,n}$                &$C^{r}_{z,n}$                            &$r_{z,t,n}$                  \\
                storages (power)    &$\widetilde{S}_{z,k}$                &$\widetilde{s}^{+}_{z,k}$        & NA &$C^{s}_{z,k}$                             &$s_{z,t,k}$\\
                storages (energy)   &$\widetilde{V}_{z,k}$                &$\widetilde{v}^{+}_{z,k}$        & NA &$C^{v}_{z,k}$                            & NA                            \\
                transmission &$\widetilde{X}_{z,zz}$               &$\widetilde{x}^{+}_{z,zz}$       & NA &$C^{x}_{z,zz}$                           &$x_{z,zz,t}$\\
                \bottomrule
            \end{tabulary}

            \begin{tablenotes}
                \item[$\dagger$] parameter
                \item[$\ddagger$] variable
            \end{tablenotes}
        \end{threeparttable}
    \end{table*}

\newpage
\subsection{Mathematical description} \label{subsec:apdx_mathmodel}

    \paragraph{Model objective}
    \emph{medea} minimizes total system cost $c$, i.e. the total cost of generating electricity and heat from
    technologies and capacities adequate to meet demand, over a large number of decision variables, essentially
    representing investment and dispatch decisions in each market zone$z$of the modelled energy systems.
    \begin{align}
        \min c =\sum_{z} (c_{z})
    \end{align}
    Zonal system costs $c_{z}$ are the sum of fuel cost $c^{b}_{z,t,i}$, emission cost $c^{e}_{z,t,i}$, operation and
    maintenance cost, capital costs of investment in conventional and intermittent generation ($c^{g}_{z}$,$c^{r}_{z}$),
    storage ($c^{s,v}_{z}$) and transmission ($c^{x}_{z}$) equipment, and the cost of non-served load ($c^{q}_{z}$) that
    accrues when demand is not met, e.g. when there is a power outage.
    We set $c^{q}_{z} = 12500$.
    \begin{align}
        c_{z} =\sum_{t,i}  c^{b}_{z,t,i} +\sum_{t,i} c^{e}_{z,t,i} +\sum_{i} c^{om}_{z,i} +\sum_{n} c^{om}_{z,n} + c^{g}_{z} +
        c^{r}_{z} + c^{s,v}_{z} + c^{x}_{z} + c^{q}_{z} \qquad \qquad \forall z
    \end{align}
    The components of zonal system costs are calculated according to \autoref{fuel_cost} to \autoref{lost_load_cost}.
    Lower-case $c$ represent total cost, while upper-case $C$ denotes the specific, annualized capital cost of technology investment.
    Prices for fuels and \ce{CO2} are denoted by $P$.
    \begin{align}
        &c^{b}_{z,t,i}& =&\ \sum_{f \in \Omega} \left( P_{t,z,f}\:b_{t,z,i,f}\right)\qquad \qquad&\forall z,t,i\label{fuel_cost}\\
        &c^{e}_{z,t,i}& =&\ \sum_{f}\left( P^{e}_{t,z}\:e_{z,t,i}\right)\qquad \qquad&\forall z,t,i\\
        &c^{om}_{z,i}& =&\ \left( \widetilde{G}_{z,i} - \widetilde{g}^{-}_{z,i} + \widetilde{g}^{+}_{z,i}\right) \Bar{O}^{g}_{i} + \sum_{t}\sum_{m}\sum_{f \in \Omega}\left( O^{g}_{i}\:g_{z,t,i,m,f}\right)\qquad \qquad&\forall z,i\\
        &c^{om}_{z,n}& =&\ \left( \widetilde{R}_{z,n} - \widetilde{r}^{-}_{z,n} + \widetilde{r}^{+}_{z,n}\right) \Bar{O}^{r}_{z,n} + \sum_{t}\left( O^{r}_{n}\:r_{z,t,n}\right)\qquad \qquad&\forall z,n\\
        &c^{g}_{z}& =&\ \sum_{i}\left( C^{g}_{z,i}\: \widetilde{g}^{+}_{z,i}\right)\qquad \qquad&\forall z\\
        &c^{r}_{z}& =&\ \sum_{n}\left( C^{r}_{z,n}\: \widetilde{r}^{+}_{z,n}\right)\qquad \qquad&\forall z\\
        &c^{s,v}_{z}& =&\ \sum_{k}\left( C^{s}_{z,k}\: \widetilde{s}^{+}_{z,k} + C^{v}_{z,k} \:v^{+}_{z,k}\right)\qquad \qquad&\forall z\\
        &c^{x}_{z}& =&\ \frac{1}{2}\: \sum_{zz} (C^{x}\: \delta_{z,zz}\: \widetilde{x}^{+}_{z,zz})\qquad \qquad&\forall z\label{transmission_expansion_cost}\\
        &c^{q}_{z}& =&\ \mu \sum_{t}\sum_{m} q^{-}_{z,t,m}\qquad \qquad&\forall z\label{lost_load_cost}
    \end{align}

    \paragraph{Market clearing}
    In each hour, the markets for electricity and heat have to clear.
    \autoref{market_clearing_el} ensures that the total supply from conventional and intermittent sources, and
    storages equals total electricity demand plus net exports, electricity stored and used for heat generation.
    Likewise, \autoref{market_clearing_ht} clears the heat market by equating heat generation to heat demand.
    \begin{align}
        \begin{split}
            \sum_{i}\sum_{f \in \Omega_{i,f}} g_{z,t,i,\text{el},f} + \sum_{n} r_{z,t,n} + \sum_{k} s^{out}_{z,t,k} &= \\D_{z,t,\text{el}} + \sum_{i} b_{z,t,i,\text{el}} + & \sum_{k} s^{in}_{z,t,k} + \sum_{zz} x_{z,zz,t} - q^{-}_{z,t,\text{el}} + q^{+}_{z,t} \qquad \forall z,t
        \end{split}
        \label{market_clearing_el}
    \end{align}
    \begin{align}
        \sum_{i}\sum_{f \in \Omega_{i,f}} g_{z,t,i,\text{ht},f} = D_{z,t,\text{ht}} - q^{-}_{z,t,\text{ht}}\qquad \forall z,t\label{market_clearing_ht}
    \end{align}
    \emph{medea} can be thought of as representing energy-only electricity and heat markets without capacity payments.
    Then, the marginals of the market clearing equations (\ref{market_clearing_el}) and (\ref{market_clearing_ht}),
    $\partial C / \partial D_{z,t,m}$, can be interpreted as the zonal prices for electricity and heat, respectively.

    \paragraph{Energy generation}
    Energy generation $g_{z,t,i,m,f} \geq 0$ is constrained by available installed capacity, which can be adjusted
    through investment ($\widetilde{g}^{+}_{z,i} \geq 0$) and decommissioning $\widetilde{g}^{-}_{z,i} \geq 0$.
    \begin{align}
        \sum_{f \in  \Omega} g_{z,t,i,m,f}\leq \widetilde{G}_{z,i} +\widetilde{g}^{+}_{z,i} -\widetilde{g}^{-}_{z,i}\qquad \qquad \forall z,t,i,m
    \end{align}
    Generator efficiency $\eta$ determines the amount of fuel $b_{z,t,i,f} \geq 0$ that needs to be spent in order to generate a given amount of energy.
    \begin{align}
        g_{z,t,i,m,f} = \sum_{f}\eta_{i,m,f}\:&b_{z,t,i,f}\qquad \qquad \forall z,t,i \notin J, f \\
        &b_{z,t,i,f \notin \Omega_{i,f}} = 0
    \end{align}

    \paragraph{Thermal co-generation}
    Co-generation units jointly generate heat and electricity. All feasible combinations of heat and electricity
    generation along with the corresponding fuel requirement are reflected in so-called `feasible operating regions'.
    The elements $l \in L$ span up a three-dimensional, convex feasible operating region for each co-generation technology.
    The weights $w_{z,t,i,l,f} \geq 0$ form a convex combination of the corners $l$, which are scaled to the available
    installed capacity of each co-generation technology.
    Defining weights over fuels allows co-generation units to switch fuels between multiple alternatives.
    Heat and electricity output along with the corresponding fuel requirement is then set according to the chosen weights.
    \begin{align}
        \sum_{l}\sum_{f \in \Omega} w_{z,t,i,l,f} \leq \widetilde{G}_{z,i} +\widetilde{g}^{+}_{z,i} -\widetilde{g}^{-}_{z,i}\qquad \qquad \forall z,t,i \in J\\
        g_{z,t,i,m,f} =\sum_{l}\psi_{i,l,m}\:w_{z,t,i,l,f}\qquad \qquad \forall z,t,i \in J, m, f\\
        b_{z,t,i,f} =\sum_{l}\chi_{i,l,f}\:w_{z,t,i,l,f}\qquad \qquad \forall z,t,i \in J, f\\
        w(z,t,i,l,f) = 0\qquad \qquad \forall z,t,i,k,f:\chi_{i,l,f} = 0
    \end{align}

    \paragraph{Intermittent electricity generation}
    Electricity generation from intermittent sources wind (on-shore and off-shore), solar irradiation, and river runoff
    follows generation profiles $\phi_{z,t,n} \in [0,1]$ and is scaled according to corresponding installed
    ($ \widetilde{R}_{z,n}$) and added ($\widetilde{r}^{+}_{z,n} \geq 0$) capacity.
    \begin{align}
        r_{z,t,n} = \phi_{z,t,n}\: \left( \widetilde{R}_{z,n} -\widetilde{r}^{-}_{z,n} +\widetilde{r}^{+}_{z,n}\right)\qquad \qquad \forall z,t,n
    \end{align}

    \paragraph{Electricity storages}
    Charging ($s^{in}_{z,t,k} \geq 0$) and discharging ($s^{out}_{z,t,k} \geq 0$) of storages is constrained by the
    storages' installed ($\widetilde{S}^{in}_{z,k}, \widetilde{S}^{out}_{z,k}$) and added
    ($\widetilde{s}^{+}_{z,k} \geq 0$) charging and discharging power, respectively.
    Similarly, the total energy that can be stored is constrained by the storage technology's initial
    ($\widetilde{V}_{z,k}$) and added ($\widetilde{v}^{+}_{z,k} \geq 0$) energy capacity.
    \begin{align}
        s^{out}_{z,t,k} \leq \widetilde{S}^{out}_{z,k} + \widetilde{s}^{+}_{z,k}\qquad \qquad \forall z,t,k\\
        s^{in}_{z,t,k}\leq \widetilde{S}^{in}_{z,k} + \widetilde{s}^{+}_{z,k}\qquad \qquad \forall z,t,k\\
        v_{z,t,k}\leq \widetilde{V}_{z,k} +\widetilde{v}^{+}_{z,k}\qquad \qquad \forall z,t,k
    \end{align}
    Storage operation is subject to a storage balance, such that the current energy content must be equal to the
    previous period's energy content plus all energy flowing into the storage less all energy flowing out of the storage.
    \begin{align}
        v_{z,t,k} = \rho_{z,t,k} + \eta^{in}_{z,k}\:s^{in}_{z,t,k} - (\eta^{out}_{z,k})^{-1}\:s^{out}_{z,t,k} + v_{z,t-1,k}\qquad \qquad \forall z,t,k: t>1,\: \eta^{out}_{z,k} > 0
    \end{align}
    Since the model can add storage power capacity and energy capacity independently, we require a storage to hold at least as much energy as it could store in (or out) in one hour. 
    \begin{align}
        \widetilde{v}^{+}_{z,k}\geq \widetilde{s}^{+}_{z,k}\qquad \qquad \forall z,k
    \end{align}
    To avoid last-round effects, i.e. a complete draw-down of the storage at year end, we require the storage to be filled to the same level at the beginning ($t=1$) and the end of the year ($t=T$).
    \begin{align}
        v_{z,1,k} \geq \Bar{v} \qquad \forall z, k\\
        v_{z,T,k} \geq \Bar{v} \qquad \forall z, k
    \end{align}
    
    \paragraph{Emission accounting}
    Burning fossil fuels for energy generation produces emissions of carbon dioxide (\ce{CO2}). The amount of \ce{CO2} emitted
    is tracked by the following equation
    \begin{align}
        e_{z,t,i} = \sum_{f}\left(\varepsilon_{f}\:b_{z,t,i,f}\right)\qquad \qquad \forall z,t,i
    \end{align}
    
    \paragraph{Air pollution}
    Air pollutants are released during the manufacturing of plants and when fuel is burned. The following accounting equation keeps track of these external cost, which do not enter the objective function.
    \begin{align}
    \begin{split}
        c^{a}_{z,f} =& 
        \sum_{i} \Bar{\alpha}_{f} \left(\widetilde{G}_{z,i} 
        +\widetilde{g}^{+}_{z,i} -\widetilde{g}^{-}_{z,i}\right) \Omega_{i,f}
        + \sum_{n} \Bar{\alpha}_{f} \left( \widetilde{R}_{z,n} -\widetilde{r}^{-}_{z,n} +\widetilde{r}^{+}_{z,n}\right) \Omega_{n,f}
        \\
        + &\sum_{i} \alpha_{f} b_{z,t,i,f}
        + \sum_{n} \alpha_{f} r_{z,t,n} \Omega_{n,f}
    \end{split}
    \end{align}
    
    \paragraph{Electricity exchange}
    Implicitly, \emph{medea} assumes that there are no transmission constraints within market zones.
    However, electricity exchange between market zones is subject to several constraints.

    First, exchange between market zones is constrained by available transfer capacities. Transfer capacities can be
    expanded at constant, specific investment cost (see \autoref{transmission_expansion_cost}).
    This rules out economies of scale in transmission investment that might arise in interconnected, meshed grids.
    \begin{align}
        x_{z,zz,t} \leq \widetilde{X}_{z,zz} + \widetilde{x}^{+}_{z,zz}\qquad \qquad \forall z, zz, t\\
        x_{z,zz,t}\geq-\left( \widetilde{X}_{z,zz} +\widetilde{x}^{+}_{z,zz}\right)\qquad \qquad \forall z, zz, t
    \end{align}
    By definition, electricity net exports $x_{z,zz,t}$ from $z$ to $zz$ must equal electricity net imports of $zz$ from $z$.
    \begin{align}
        x_{z,zz,t} = -x_{zz,z,t} \qquad \qquad \forall z, zz, t
    \end{align}
    Added transmission capacities can be used in either direction.
    \begin{align}
        \widetilde{x}^{+}_{z,zz} = \widetilde{x}^{+}_{zz,z}\qquad \qquad \forall z, zz
    \end{align}
    Finally, electricity cannot flow between zones where there is no transmission infrastructure in place (including
    intra-zonal flows).
    \begin{align}
        x_{z,zz,t} = 0 \qquad \qquad \forall z, zz, t:\widetilde{X}_{z,zz} = 0\\
        x_{zz,z,t} = 0\qquad \qquad \forall z, zz, t:\widetilde{X}_{z,zz} = 0
    \end{align}

    \paragraph{Decommissioning}
    Keeping plant available for generation gives rise to quasi-fixed operation and maintenance costs.
    Such cost can be avoided by decommissioning an energy generator. This is modelled as a reduction in generation
    capacity, which cannot exceed installed capacity.
    \begin{align}
        \widetilde{g}^{-}_{z,i} &\leq \widetilde{G}_{z,i} + \widetilde{g}^{+}_{z,i}\qquad \qquad \forall z,i\\
        \widetilde{r}^{-}_{z,n} &\leq \widetilde{R}_{z,n} +\widetilde{r}^{+}_{z,n}\qquad \qquad \forall z,n
    \end{align}

    \paragraph{Ancillary services}
    Power systems require various system services for secure and reliable operation, such as balancing services or
    voltage support through the provision of reactive power. Such system services can only be supplied by operational
    generators.
    Thus, we approximate system service provision by a requirement on the minimal amount of spinning reserves operating
    at each hour.
    We assume that ancillary services are supplied by conventional (thermal) power plant, hydro power plant, and storage.
    The requirement for spinning reserves is assumed to be proportional to electricity peak load
    $\widehat{D}_{z,\text{el}} = \max_{t} D_{z,t,\text{el}}$ and peak generation from wind and solar resources,
    where $\widehat{\phi}_{z,n} = \max_{t} \phi_{z,t,n}$.
    \begin{align}
        \sum_{i}\sum_{f \in \Omega}\left( g_{z,t,i,\text{el},f}\right) + r_{z,t,\text{ror}}
            + \sum_{k}\left( s^{out}_{z,t,k} + s^{in}_{z,t,k}\right)\geq \lambda_{z}\widehat{D}_{z,\text{el}}
            + \sigma_{z}\sum_{n\setminus \{ \text{ror}\}}\widehat{\phi}_{z,n} (\widetilde{R}_{z,n}
            +\widetilde{r}^{+}_{z,n})\qquad \forall z,t
    \end{align}

    \paragraph{Curtailment}
    Electricity generated from intermittent sources can be curtailed (disposed of) without any further cost (apart from
    implicit opportunity cost).
    \begin{align}
        q^{+}_{z,t} \leq \sum_{n\setminus \{ \text{ror}\}} r_{z,t,n} \qquad \qquad \forall z, t
    \end{align}
\end{document}